\documentclass[letterpaper,twocolumn,english,aps,pre,showpacs]{revtex4}
\usepackage{CJK}
\usepackage[latin9]{inputenc}
\usepackage{babel}
\usepackage{verbatim}
\usepackage{amsmath}
\usepackage{mathtools}
\usepackage{amssymb}
\usepackage{graphicx}
\usepackage{esint}
\usepackage[unicode=true,
 bookmarks=true,bookmarksnumbered=false,bookmarksopen=false,
 breaklinks=false,pdfborder={0 0 1},backref=false,colorlinks=false]
 {hyperref}
\hypersetup{pdftitle={Fluctuation-response RelationUnifies Dynamical Behaviors in Neural Fields}}
\usepackage{times}

\makeatletter

\@ifundefined{textcolor}{}
{%
 \definecolor{BLACK}{gray}{0}
 \definecolor{WHITE}{gray}{1}
 \definecolor{RED}{rgb}{1,0,0}
 \definecolor{GREEN}{rgb}{0,1,0}
 \definecolor{BLUE}{rgb}{0,0,1}
 \definecolor{CYAN}{cmyk}{1,0,0,0}
 \definecolor{MAGENTA}{cmyk}{0,1,0,0}
 \definecolor{YELLOW}{cmyk}{0,0,1,0}
}

\usepackage{babel}
\usepackage{babel}

\makeatother

\begin{document}

\begin{CJK*}{UTF8}{min} 

\title{Fluctuation-response Relation Unifies Dynamical Behaviors in Neural
Fields}

\author{C. C. Alan Fung (馮志聰)\textsuperscript{1}, K. Y. Michael Wong (王國彝)\textsuperscript{1},
Hongzi Mao (毛宏自)\textsuperscript{1} and Si Wu (吴思)\textsuperscript{2}}

\affiliation{\textsuperscript{1}Department of Physics, The Hong Kong University
of Science and Technology, Clear Water Bay, Hong Kong, China}

\affiliation{\textsuperscript{2}State Key Laboratory of Cognitive Neuroscience
and Learning, IDG/McGovern Institute for Brain Research, Beijing Normal
University, Beijing 100875, China }
\begin{abstract}
Anticipation is a strategy used by neural fields to compensate for
transmission and processing delays during the tracking of dynamical
information, and can be achieved by slow, localized, inhibitory feedback
mechanisms such as short-term synaptic depression, spike-frequency
adaptation, or inhibitory feedback from other layers. Based on the
translational symmetry of the mobile network states, we derive generic
fluctuation-response relations, providing unified predictions that
link their tracking behaviors in the presence of external stimuli
to the intrinsic dynamics of the neural fields in their absence. 
\end{abstract}


\maketitle

\end{CJK*}




\section{Introduction}

It is well known that there is a close relation between the fluctuation
properties of a system near equilibrium and its response to external
driving fields. Brownian particles diffusing rapidly when left alone
have a high mobility when driven by external forces (Einstein\textendash Smoluchowski
Relation) \cite{Einstein1905,vonSmoluchowski1906}. Electrical conductors
with large Johnson\textendash Nyquist noise have high conductivities
\cite{Nyquist1928}. Materials with large thermal noise have low specific
heat \cite{Huang1987}. These fluctuation-response relations (FRRs)
unify the intrinsic and extrinsic properties of many physical systems.

Fluctuations are relevant to neural systems processing continuous
information such as orientation \cite{Ben-Yishai1995}, head direction
\cite{Zhang1996}, and spatial location \cite{Samsonovich1997}. It
is commonly believed that these systems represent external information
by localized activity profiles in neural substrates, commonly known
as neural fields \cite{Wilson1972,Amari1977}. Analogous to particle
diffusion, location fluctuations of these states represent distortions
of the information they represent, and at the same time indicate their
mobility under external influences. When the motion of these states
represents moving stimuli, their mobility will determine their responses,
such as the amount of time delay when they track moving stimuli. This
provides the context for the application of the FRR.

In processing time-dependent external information, real-time response
is an important and even a life-and-death issue to animals. However,
time delay is pervasive in the dynamics of neural systems. For example,
it takes 50 -- 80 ms for electrical signals to transmit from the retina
to the primary visual cortex \cite{Nijhawan2009}, and 10 -- 20 ms
for a neuron to process and integrate temporal input in such tasks
as speech recognition and motor control.

To achieve real-time tracking of moving stimuli, a way to compensate
delays is to predict their future position. This is evident in experiments
on the head-direction (HD) systems of rodents during head movements
\cite{O'Keefe1971,Taube1990}, in which the direction perceived by
the HD neurons has nearly zero lag with respect to the true instantaneous
position \cite{Taube1998}, or can even lead the current position
by a constant time \cite{Blair1995}. This anticipative behavior is
also observed when animals make saccadic eye movements \cite{Sommer2006}.
In psychophysics experiments, the future position of a continuously
moving object is anticipated, but intermittent flashes are not \cite{Nijhawan1994}.

There are different delay compensation strategies, and many of them
have slow, local inhibitory feedback in their dynamics. For example,
short-term synaptic depression (STD) can implement anticipatory tracking
\cite{Fung2013}. Its underlying mechanism is the slow depletion of
neurotransmitters in the active region of the network state, facilitating
neural fields to exhibit a rich spectrum of dynamical behaviors \cite{Wang2015}.
This depletion increases the tendency of the network state to shift
to neighboring positions. For sufficiently strong STD, the tracking
state can even overtake the moving stimulus. At the same time, local
inhibitory feedbacks can induce spontaneous motion of the localized
states in neural fields \cite{Ben-Yishai1997,York2009,Fung2012}.
Remarkably, the parameter region of anticipatory tracking is effectively
identical to that of spontaneous motion. Since spontaneous motion
sets in when location fluctuation diverges, this indicates the close
relation between fluctuations and responses, and implies that such
a relation should be more generic than the STD mechanism itself.

Besides STD, other mechanisms can also provide slow, local inhibitory
feedback to neurons. Examples include spike-frequency adaptation (SFA)
that refers to the reduction of neuron excitability after prolonged
stimulation \cite{Treves1993}, and inhibitory feedback loops (IFL)
in multilayer networks that refer to the negative feedback interaction
via feedback synapses from the downstream neurons \cite{Zhang2012}
in both one dimension and two dimensions \cite{Fung2015}. Like STD,
such local inhibition can generate spontaneous traveling waves \cite{Ben-Yishai1997}.
Likewise, they are expected to exhibit anticipatory tracking \cite{Zhang2012}.
In this paper, we will consider how FRR provides a unified picture
for this family of systems driven by different neural mechanisms.
As will be shown, generic analyses based on the translational symmetry
of the systems show that anticipative tracking is associated with
spontaneous motions, thus providing a natural mechanism for delay
compensation.

\section{General Mathematical Framework of Neural Field Models}

We consider a neural field in which neurons are characterized by location
$x$, interpreted as the preferred stimulus of the neuron, which can
be spatial location \cite{Samsonovich1997} or head direction \cite{Zhang1996}.
Neuronal activities are represented by $u(x,t)$, interpreted as neuronal
current \cite{Wu2008,Fung2010}. To keep the formulation generic,
the dynamical equation is written in the form 
\begin{equation}
\frac{\partial u\left(x,t\right)}{\partial t}=F_{u}\left[x;u,p\right]+I^{{\rm ext}}(x,t).\label{eq:dudt}
\end{equation}
$F_{u}$ is a functional of $u$ and $p$ evaluated at $x$. $p$
is a dynamical variable representing neuronal activities with no direct
connections with the external environment. In the context of anticipatory
tracking, $p$ corresponds to a dynamical local inhibitory mechanism.
It could represent the available amount of neurotransmitters of presynaptic
neurons for STD \cite{Tsodyks1997,Fung2012}, or the shift of the
firing thresholds due to SFA \cite{Treves1993}, or the neuronal activities
of a hidden neural field layer in IFL \cite{Zhang2012}. Explicit
forms of $F_{u}\left[x;u,p\right]$ for STD, SFA and IFL can be found
in the next section. Besides the force $F_{u}$, the dynamics is also
driven by an external input, $I^{{\rm ext}}$.

Similar to Eq. (\ref{eq:dudt}), the dynamics of $p$ is given by
\begin{equation}
\frac{\partial p\left(x,t\right)}{\partial t}=F_{p}\left[x;u,p\right].\label{eq:dpdt}
\end{equation}
$F_{p}$ is also a functional of $u$ and $p$ evaluated at $x$.
Explicit expressions of $F_{p}$ for STD, SFA and IFL can also be
found in the next section. For the present analysis, it is sufficient
to assume that (i) the forces are translationally invariant, and (ii)
the forces possess inversion symmetry.

\section{Example Models}

The formalism we quoted in the previous section is generic. To test
the general results deduced from the generic formalism, we have chosen
three models with different kinds of dynamical local inhibitory mechanisms.
They are spike frequency adaptation (SFA), short-term synaptic depression
(STD) and inhibitory feedback loop (IFL). All these models are based
on the model proposed by Wu \textit{et al}. \cite{Wu2008} and studied
in detail by Fung \textit{et al}. \cite{Fung2010}. However, the studied
behaviors are applicable to general models.

\subsection{Neural Field Model with Spike Frequency Adaptation}

For spike frequency adaptation (SFA), $F_{u}$ is given by \cite{Mi2014}
\begin{align}
F_{u}\left[x;u,p\right]\equiv\frac{1}{\tau_{{\rm s}}}\Biggl[ & \rho\int dx^{\prime}J\left(x,x^{\prime}\right)r\left(x^{\prime},t\right)\Biggr.\nonumber \\
 & \Biggl.-p\left(x,t\right)-u\left(x,t\right)\Biggr].\label{eq:Fu_sfa}
\end{align}
$\tau_{s}$ is the timescale of $u(x,t)$, which is of the order of
the magnitude of 1 ms. For simplicity, neurons in the preferred stimulus
space are distributed evenly. $\rho$ is the density of neurons in
the preferred stimulus space. $J(x,x^{\prime})$ is the excitatory
coupling between neurons at $x$ and $x^{\prime}$, which is given
by 
\begin{equation}
J\left(x,x^{\prime}\right)\equiv\frac{J_{0}}{\sqrt{2\pi}a}\exp\left(\frac{\left|x-x^{\prime}\right|^{2}}{2a^{2}}\right).\label{eq:Jxx}
\end{equation}
This coupling depends only on the difference between the preferred
stimuli of neurons. So this coupling function is translationally invariant.
Here, $a$ is the range of the excitatory coupling in the space, while
$J_{0}$ is the strength of the excitatory coupling. $r(x,t)$ is
the neuronal activity of neurons at $x$. It depends on $u(x,t)$.
We define it to be 
\begin{equation}
r\left(x,t\right)\equiv\frac{\max\left[u\left(x,t\right),0\right]{}^{2}}{1+k\rho\int dx^{\prime}\max\left[u\left(x^{\prime},t\right),0\right]^{2}},
\end{equation}
where $k$ is the global inhibition. The integral in Eq. (\ref{eq:Fu_sfa})
is the weighted sum of the excitatory signal from different neurons
in the neuronal network.

On the right hand side of Eq. (\ref{eq:Fu_sfa}), $-u(x,t)$ is the
relaxation, while $p(x,t)$ is the dynamical variable modelling the
effect of SFA. Its dynamics is defined by \cite{Mi2014} 
\begin{equation}
F_{p}\left[x;u,p\right]\equiv\frac{1}{\tau_{i}}\Biggl\{-p\left(x,t\right)+\gamma\max\left[u\left(x,t\right),0\right]\Biggr\}.\label{eq:Fp_sfa}
\end{equation}
$\tau_{i}$ is the time scale of $p(x,t)$, which is of the order
of 100 ms. $\gamma$ is the strength of SFA.

In Eq. (\ref{eq:dudt}), $I^{{\rm ext}}(x,t)$ is the external input.
For convenience, it is chosen to be 
\begin{equation}
I^{{\rm ext}}\left(x,t\right)\equiv\frac{A}{\tau_{s}}\exp\left[-\frac{\left|x-z_{I}\left(t\right)\right|^{2}}{4a^{2}}\right].\label{eq:Iext}
\end{equation}
$A$ is the magnitude of the external input, while $z_{I}$ is the
position of the external input. Note that the exact choice should
not alter our conclusion in the weak external input limit \cite{Fung2010}.

\subsection{Neural Field Model with Short-term Synaptic Depression}

For short-term synaptic depression (STD), $F_{u}$ is defined by 
\begin{align}
F_{u}\left[x;u,p\right]\equiv\frac{1}{\tau_{{\rm s}}}\Biggl[ & \rho\int dx^{\prime}J\left(x,x^{\prime}\right)p\left(x^{\prime},t\right)r\left(x^{\prime},t\right)\Biggr.\nonumber \\
 & \Biggl.-u\left(x,t\right)\Biggr].\label{eq:Fu_std}
\end{align}
Notations are the same as those in Eq. (\ref{eq:Fu_sfa}), except
that $p(x,t)$ models the multiplicative effect due to STD \cite{Fung2012}.
Here the physical meaning of $p(x,t)$ is the available portion of
neurotransmitters in the presynaptic neurons with preferred stimulus
$x$ at time $t$.

The dynamics of $p(x,t)$ is given by \cite{Tsodyks1997,Fung2012}
\begin{equation}
F_{p}\left[x;u,p\right]\equiv\frac{1}{\tau_{d}}\left[1-p\left(x,t\right)-\tau_{d}\beta p\left(x,t\right)r\left(x,t\right)\right].\label{eq:Fp_std}
\end{equation}
$\tau_{d}$ is the time scale of STD, which is of the order of 100
ms. $\beta$ is the strength of STD.

\subsection{Neural Field Model with an Inhibitory Feedback Loop}

For neural field models with an inhibitory feedback loop (IFL), \cite{Zhang2012}
\begin{align}
F_{u}\left[x;u,p\right] & \equiv\frac{1}{\tau_{1}}\Biggl[-u\left(x,t\right)+\rho\int dx^{\prime}J\left(x,x^{\prime}\right)r_{u}\left(x^{\prime},t\right)\Biggr.\nonumber \\
 & \qquad\quad\Biggl.+\left(\frac{J_{{\rm fb}}}{J_{0}}\right)\rho\int dx^{\prime}J\left(x,x^{\prime}\right)r_{p}\left(x^{\prime},t\right)\Biggr],\label{eq:Fu_ifl}\\
F_{p}\left[x;u,p\right] & \equiv\frac{1}{\tau_{2}}\Biggl[-p\left(x,t\right)+\rho\int dx^{\prime}J\left(x,x^{\prime}\right)r_{p}\left(x^{\prime},t\right)\Biggr.\nonumber \\
 & \qquad\quad\Biggl.+\left(\frac{J_{{\rm ff}}}{J_{0}}\right)\rho\int dx^{\prime}J\left(x,x^{\prime}\right)r_{u}\left(x^{\prime},t\right)\Biggr].\label{eq:Fp_ifl}
\end{align}
So is Eq. (\ref{eq:Fu_std}). Notations are the same as those in Eq.
(\ref{eq:Fu_sfa}), except that $p(x,t)$ is the network state of
the inhibitory feedback loop. $r_{u/p}$ are defined by 
\begin{equation}
r_{i}\left(x,t\right)\equiv\frac{\max\left[u_{i}\left(x,t\right),0\right]{}^{2}}{1+k\rho\int dx^{\prime}\max\left[u_{i}\left(x^{\prime},t\right),0\right]^{2}},
\end{equation}
where $i$ is $u$ or $p$.

$J_{{\rm ff}}$ is the strength of the feedforward connection from
the $u$-layer to the $p$-layer, while $J_{{\rm fb}}$ is the strength
of the feedback connection from the $p$-layer to the $u$-layer.
$\tau_{1}=\tau_{s}$ and $\tau_{2}$ are the time scales of $u(x,t)$
and $p(x,t)$ respectively. They are of the order of 1 ms. In this
work, for simplicity, we assume them to be the same. However, as shown
in Appendix \ref{sec:IFL}, the slowness of the inhibitory feedback
arises from the weak coupling between the exposed and inhibitory layers.

\subsection{Rescaling of Parameters and Variables}

It is convenient to present results and choice of parameters in the
rescaled manner. Following the rescaling rules in \cite{Fung2012},
we define $\tilde{u}(x,t)\equiv\rho J_{0}u(x,t)$ and $\tilde{A}\equiv\rho J_{0}A$.
For SFA, since $p$ has a same dimension as $u$, we define $\tilde{p}$
in the same way as $u$: $\tilde{p}\equiv\rho J_{0}p$. For STD, $p(x,t)$
is dimensionless, and we rescale $\beta$ according to $\tilde{\beta}\equiv\tau_{d}\beta/(\rho^{2}J_{0}{}^{2})$.
For IFL, we rescale $u$ and $p$ in the same way we have done for
SFA. For our convenience, we define $\tilde{J}_{{\rm ff}}\equiv J_{{\rm ff}}/J_{0}$
and $\tilde{J}_{{\rm fb}}\equiv J_{{\rm fb}}/J_{0}$. In these three
cases, we need to rescale $k$ as well. As in \cite{Fung2010}, for
$\beta=0$, $\gamma=0$ and $J_{{\rm fb}}=0$, the stable steady state
exists only when $k<k_{c}\equiv\rho J_{0}{}^{2}/(8\sqrt{2\pi}a)$.
Hence we define $\tilde{k}\equiv k/k_{c}$ to simplify our presentation
of parameters.


\begin{figure*}[t]
\begin{centering}
\includegraphics[width=1\textwidth]{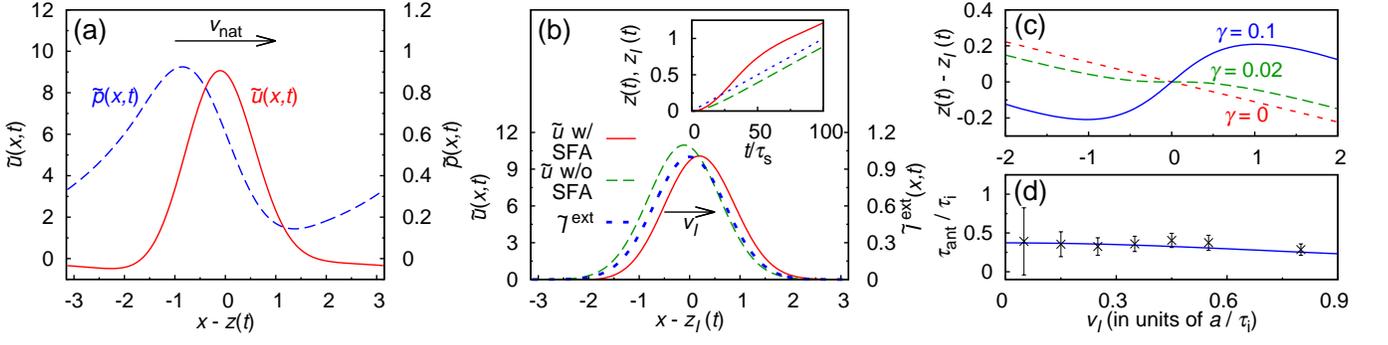} 
\par\end{centering}

\protect\protect\caption{\label{fig:moving_bump} (color online) (a) The rescaled neuronal
current, $\tilde{u}(x,t)$, and the rescaled inhibitory variable for
SFA, $\tilde{p}(x,t)$ during a spontaneous motion in the moving frame
centered at $z(t)$. $z(t)$ is the center of mass of $\tilde{u}(x,t)$.
The $\tilde{u}(x,t)$ profile is moving to the direction pointed by
the arrow. Parameters: $\tilde{k}$ (rescaled inhibition) = 0.5, $\gamma$
(SFA strength) = 0.2, $\tau_{{\rm s}}$ (time constant of neuronal
current) = 1 ms and $\tau_{{\rm i}}$ (time constant of SFA) = 50
ms. (b) $\tilde{u}(x,t)$ and $\tilde{I}^{\text{ext}}(x,t)$, rescaled
external stimulus, during a tracking process. Inset: $z_{0}(t)$ and
$z(t)$, the centers of mass of $\tilde{I}^{\text{ext}}(x,t)$ and
$\tilde{u}(x,t)$, respectively. The $\tilde{I}^{\text{ext}}(x,t)$
profile is moving in the direction of the arrow with velocity $v_{I}$.
Parameters: $\tilde{k}=0.5$, $\gamma=0$, $\tau_{{\rm s}}=1\text{ ms}$,
$\tilde{A}$ (rescaled magnitude of $\tilde{I}^{{\rm ext}}$) = 1.0
and $v_{I}=0.01$. (c) Displacement of the $\tilde{u}$ profile relative
to the external stimulus, $z(t)-z_{0}(t)$. Parameters: $\tilde{k}=0.5$,
$\tau_{{\rm i}}=50\text{ ms}$ and $\tau_{{\rm s}}=1\text{ ms}$.
(d) Curve: The anticipation time, $\tau_{\text{ant}}\equiv\left[z(t)-z_{0}(t)\right]/v_{I}$,
for the case with $\gamma=0.1$ in (c). Symbols: Anticipation time
in Fig. 4 of \cite{Goodridge2000} with the assumption that $\tau_{i}=50$~ms
and $a=22.5^{\circ}$.}
\end{figure*}

\section{Translational Invariance and Inversion Symmetry}

Studies on neural field models showed that they can support a profile
of localized activities even in the absence of external stimuli \cite{Wilson1972,Amari1977,Ben-Yishai1995,Fung2010}.
Irrespective of the explicit form of this ``bump'', it is sufficient
to note that there exists a non-trivial stable solution $\{u_{0},p_{0}\}$
satisfying 

\begin{equation}
F_{u}[x;u_{0},p_{0}]=F_{p}[x;u_{0},p_{0}]=0,
\end{equation}
and that this solution is neutrally stable in $x$, that is, for an
arbitrary bump position $z$,

\begin{equation}
F_{u}[x-z;u_{0},p_{0}]=F_{p}[x-z;u_{0},p_{0}]=0.
\end{equation}

To study the stability issue of stationary state $(u_{0},p_{0})$,
we consider the dynamics of the fluctuations about the steady state,
\begin{align}
\frac{\partial}{\partial t}\delta u\left(x\right) & =\int dx^{\prime}\frac{\partial F_{u}\left(x\right)}{\partial u\left(x^{\prime}\right)}\delta u\left(x^{\prime}\right)+\int dx^{\prime}\frac{\partial F_{u}\left(x\right)}{\partial p\left(x^{\prime}\right)}\delta p\left(x^{\prime}\right),\\
\frac{\partial}{\partial t}\delta p\left(x\right) & =\int dx^{\prime}\frac{\partial F_{p}\left(x\right)}{\partial u\left(x^{\prime}\right)}\delta u\left(x^{\prime}\right)+\int dx^{\prime}\frac{\partial F_{p}\left(x\right)}{\partial p\left(x^{\prime}\right)}\delta p\left(x^{\prime}\right).
\end{align}
Here $\delta u(x)\equiv u(x)-u_{0}(x)$ and $\delta p(x)\equiv p(x)-p_{0}(x)$.
Consider the solutions of these equations with time dependence $\exp(-\lambda t)$.
Then the eigenvalue equations become the $\Delta x\rightarrow0$ limit
of the matrix eigenvalue equation 
\begin{align}
 & \left(\begin{array}{cc}
\left\{ \frac{\partial F_{u}\left(x_{i}\right)}{\partial u\left(x_{j}\right)}\right\}  & \left\{ \frac{\partial F_{u}\left(x_{i}\right)}{\partial p\left(x_{j}\right)}\right\} \\
\left\{ \frac{\partial F_{p}\left(x_{i}\right)}{\partial u\left(x_{j}\right)}\right\}  & \left\{ \frac{\partial F_{p}\left(x_{i}\right)}{\partial p\left(x_{j}\right)}\right\} 
\end{array}\right)\left(\begin{array}{c}
\left\{ f_{u}\left(x_{j}\right)\right\} \\
\left\{ f_{p}\left(x_{j}\right)\right\} 
\end{array}\right)\Delta x\nonumber \\
= & -\lambda\left(\begin{array}{c}
\left\{ f_{u}\left(x_{i}\right)\right\} \\
\left\{ f_{p}\left(x_{i}\right)\right\} 
\end{array}\right).
\end{align}
The left eigenvector with the same eigenvalue is given by 
\begin{align}
 & \left(\begin{array}{cc}
\left\{ g_{u}\left(x_{j}\right)\right\}  & \left\{ g_{p}\left(x_{j}\right)\right\} \end{array}\right)\left(\begin{array}{cc}
\left\{ \frac{\partial F_{u}\left(x_{i}\right)}{\partial u\left(x_{j}\right)}\right\}  & \left\{ \frac{\partial F_{u}\left(x_{i}\right)}{\partial p\left(x_{j}\right)}\right\} \\
\left\{ \frac{\partial F_{p}\left(x_{i}\right)}{\partial u\left(x_{j}\right)}\right\}  & \left\{ \frac{\partial F_{p}\left(x_{i}\right)}{\partial p\left(x_{j}\right)}\right\} 
\end{array}\right)\Delta x\nonumber \\
= & -\lambda\left(\begin{array}{cc}
\left\{ g_{u}\left(x_{i}\right)\right\}  & \left\{ g_{p}\left(x_{i}\right)\right\} \end{array}\right).
\end{align}
Translational invariance implies that $\partial u_{0}/\partial x$
and $\partial p_{0}/\partial x$ are the components of the right eigenfunction
of the dynamical equations with eigenvalue 0, satisfying 
\begin{align}
\int dx^{\prime}\frac{\partial F_{u}\left(x\right)}{\partial u\left(x^{\prime}\right)}\frac{\partial u_{0}\left(x^{\prime}\right)}{\partial x^{\prime}}+\int dx^{\prime}\frac{\partial F_{u}\left(x\right)}{\partial p\left(x^{\prime}\right)}\frac{\partial p_{0}\left(x^{\prime}\right)}{\partial x^{\prime}}= & ~0,\label{eq:eigen0_ru}\\
\int dx^{\prime}\frac{\partial F_{p}\left(x\right)}{\partial u\left(x^{\prime}\right)}\frac{\partial u_{0}\left(x^{\prime}\right)}{\partial x^{\prime}}+\int dx^{\prime}\frac{\partial F_{p}\left(x\right)}{\partial p\left(x^{\prime}\right)}\frac{\partial p_{0}\left(x^{\prime}\right)}{\partial x^{\prime}}= & ~0.\label{eq:eigen0_rp}
\end{align}
The corresponding left eigenfunctions satisfy 
\begin{align}
\int dx^{\prime}g_{u}^{0}\left(x^{\prime}\right)\frac{\partial F_{u}\left(x'\right)}{\partial u(x)}+\int dx^{\prime}g_{p}^{0}\left(x^{\prime}\right)\frac{\partial F_{p}\left(x'\right)}{\partial u(x)} & =~0,\label{eq:eigen0_lu}\\
\int dx^{\prime}g_{u}^{0}\left(x^{\prime}\right)\frac{\partial F_{u}\left(x'\right)}{\partial p(x)}+\int dx^{\prime}g_{p}^{0}\left(x^{\prime}\right)\frac{\partial F_{p}\left(x'\right)}{\partial p(x)} & =~0.\label{eq:eigen0_lp}
\end{align}
For stable bumps, the eigenvalues of all other eigenfunctions are
at most 0. Let $f_{u}^{n}$ and $f{}_{p}^{n}$ be the components of
the eigenfunction with the $n^{{\rm th}}$ eigenvalue $-\lambda_{n}$,
satisfying 
\begin{align}
\int dx^{\prime}\frac{\partial F_{u}\left(x\right)}{\partial u\left(x^{\prime}\right)}f_{u}^{n}\left(x^{\prime}\right)+\int dx^{\prime}\frac{\partial F_{u}\left(x\right)}{\partial p\left(x^{\prime}\right)}f_{p}^{n}\left(x^{\prime}\right)\nonumber \\
=-\lambda_{n} & f{}_{u}^{n}\left(x\right),\\
\int dx^{\prime}\frac{\partial F_{p}\left(x\right)}{\partial u\left(x^{\prime}\right)}f{}_{u}^{n}\left(x^{\prime}\right)+\int dx^{\prime}\frac{\partial F_{p}\left(x\right)}{\partial p\left(x^{\prime}\right)}f^{n}{}_{p}\left(x^{\prime}\right)\nonumber \\
=-\lambda_{n} & f_{p}^{n}\left(x\right).
\end{align}

Similarly, denoting the components of the left eigenfunctions as $g_{u}^{n}$
and $g_{p}^{n}$ respectively,

\begin{align}
\int dx^{\prime}g_{u}^{n}\left(x^{\prime}\right)\frac{\partial F_{u}\left(x'\right)}{\partial u(x)}+\int dx^{\prime}g_{p}^{n}\left(x^{\prime}\right)\frac{\partial F_{p}\left(x'\right)}{\partial u(x)}\nonumber \\
=-\lambda_{n} & g_{u}^{n}\left(x\right),\\
\int dx^{\prime}g_{u}^{n}\left(x^{\prime}\right)\frac{\partial F_{u}\left(x'\right)}{\partial p(x)}+\int dx^{\prime}g_{p}^{n}\left(x^{\prime}\right)\frac{\partial F_{p}\left(x'\right)}{\partial p(x)}\nonumber \\
=-\lambda_{n} & g_{p}^{n}\left(x\right).
\end{align}

The eigenfunctions corresponding to eigenvalues $\lambda_{m}$ and
$\lambda_{n}$ satisfy the orthogonality condition 
\begin{equation}
\int dx^{\prime}g_{u}^{m}\left(x^{\prime}\right)f_{u}^{n}\left(x^{\prime}\right)+\int dx^{\prime}g_{p}^{m}\left(x^{\prime}\right)f_{p}^{n}\left(x^{\prime}\right)=\delta_{mn}.
\end{equation}
For later use, we define 
\begin{equation}
Q_{\psi\varphi}\equiv\int dxg_{\psi}^{0}\left(x\right)\int dx^{\prime}\frac{\partial F_{\psi}\left(x\right)}{\partial\varphi\left(x^{\prime}\right)}\frac{\partial\varphi_{0}\left(x^{\prime}\right)}{\partial x^{\prime}},
\end{equation}
where $\psi,\varphi\in\{u,p\}$. The following identities are the
results of translational invariance. Multiplying both sides of Eq.
(\ref{eq:eigen0_ru}) by $g_{u}^{0}(x)$ and integrating over $x$,
we obtain
\begin{align}
Q_{uu}+Q_{up} & =~0.
\end{align}
Similarly, multiplying both sides of Eq. (\ref{eq:eigen0_rp}) by
$g_{p}^{0}(x)$ and integrating over $x$, we have 
\begin{align}
Q_{pu}+Q_{pp} & =~0.
\end{align}
Likewise, from Eqs. (\ref{eq:eigen0_lu}) and (\ref{eq:eigen0_lp}),
we find

\begin{align}
Q_{uu}+Q_{pu}=Q_{up}+Q_{pp} & =~0.
\end{align}

Next, we consider the implications of inversion symmetry, that is,
$\partial F_{\psi}\left(x\right)/\partial\varphi\left(x^{\prime}\right)=\partial F_{\psi}\left(-x\right)/\partial\varphi\left(-x^{\prime}\right)$
for $\psi,\varphi\in\{u,p\}$. Then the dynamics preserves parity.
Suppose the bump state $u_{0}(x)$ and $p_{0}(x)$ has even parity.
Then the distortion mode $\partial u_{0}/\partial x$ and $\partial p_{0}/\partial x$
has odd parity. Note that the corresponding left eigenfunctions $g_{u}^{0}$
and $g_{p}^{0}$ have the same parity as the right eigenfunctions.

\section{Intrinsic Behavior}

Studies on neural field models with STD \cite{York2009,Fung2012},
SFA \cite{Coombes2005} and IFL \cite{Zhang2012} suggested that the
network can support spontaneously moving profiles, even though there
is no external moving input. This occurs when the static solution
becomes unstable to positional displacement in some parameter regions.
To study the stability issue of static solutions due to positional
displacement, we consider 
\begin{align}
u\left(x,t\right) & =u_{0}\left(x\right)+c_{0}\frac{\partial u_{0}\left(x\right)}{\partial x},\\
p\left(x,t\right) & =p_{0}\left(x\right)+\varepsilon_{0}\frac{\partial p_{0}\left(x\right)}{\partial x}.
\end{align}
$c_{0}$ and $\varepsilon_{0}$ are the diplacements of the exposed
and inhibitory profiles respectively (in the direction opposite to
their signs). As derived in Appendix \ref{sec:separation}, we have

\begin{equation}
\frac{d}{dt}\left(\varepsilon_{0}-c_{0}\right)=\lambda\left(\varepsilon_{0}-c_{0}\right),\label{eq:propa_stability}
\end{equation}
where the instability eigenvalue $\lambda$ is given by 
\begin{align}
\lambda & \equiv\frac{Q_{uu}}{I_{u}}+\frac{Q_{pp}}{I_{p}},\label{eq:lambda_m}
\end{align}
where $I_{\psi}=\int dxg_{\psi}^{0}(x)[d\psi_{0}(x)/dx]$ and $\psi\in\{u,p\}$.
In the static phase, where stationary solutions are stable, $\lambda<0$.
For systems with spontaneously moving bumps, $\lambda>0$. It implies
that relative displacements of stationary $u_{0}$-profile and $p_{0}$-profile
should diverge. The misalignment between the exposed $u_{0}$-profile
and hidden $p_{0}$-profile will drive the motion of $u$ to sweep
throughout the preferred stimulus space.

When the bump becomes translationally unstable, it moves with an intrinsic
speed (or natural speed). To investigate the intrinsic speed denoted
as $v_{{\rm nat}}$, we need to expand the dynamical equations beyond
first order. The small parameter is the non-vanishing profile separation
$\varepsilon_{0}$, now denoted as the intrinsic separation $\varepsilon_{{\rm int}}$.
The critical regime is given by $\varepsilon_{{\rm int}}\sim\sqrt{\lambda}$.
As derived in Appendix \ref{sec:speed},

\begin{equation}
v_{{\rm nat}}=\frac{\varepsilon_{{\rm int}}}{\tau_{{\rm int}}},\label{eq:v_nat}
\end{equation}
where

\begin{equation}
\tau_{{\rm int}}=-\frac{I_{p}}{Q_{pp}}.\label{eq:intrinsic_time}
\end{equation}

We interpret $\tau_{{\rm int}}$ as the intrinsic time scale of the
system. (We note in passing that the same result can be obtained by
substituting the moving bump solution $u(x,t)=u_{0}(x-v_{{\rm nat}}t)$,
$p(x,t)=p_{0}(x-v_{{\rm nat}}t+\varepsilon_{{\rm int}})$ into Eqs.
(\ref{eq:dudt}) and (\ref{eq:dpdt}) and expanding to the lowest
order as was done in Eq. (\ref{eq:propa_stability}). However, such
a derivation has not taken into account the stability of the solution.)

Noting that Eq. (\ref{eq:v_nat}) also holds in the static phase with
$v_{{\rm nat}}=\varepsilon_{{\rm int}}=0$, we infer that the separation
of the exposed and inhibitory profiles is the cause of the spontaneous
motion. The physical picture is that when the inhibitory profile lags
behind the exposed profile, the neuronal activity will have a stronger
tendency to shift away from the strongly inhibited region.

An example of the spontaneously moving state of neural field model
with SFA is shown in Fig. \ref{fig:moving_bump}(a), in which the
$u$-profile and $p$-profile are plotted relative to the center of
mass of $u$, $z(t)$. At the steady state of the spontaneously moving
state, the $u$-profile moves in the direction opposite to the direction
the $p$-profile biased to. So the $p$-profile always lags behind
the $u$-profile during the spontaneous motion, while $u$-profile
keeps moving due to the asymmetry granted by the misalignment between
$u$ and $p$.


\begin{figure*}[t]
\begin{centering}
\includegraphics[width=1\textwidth]{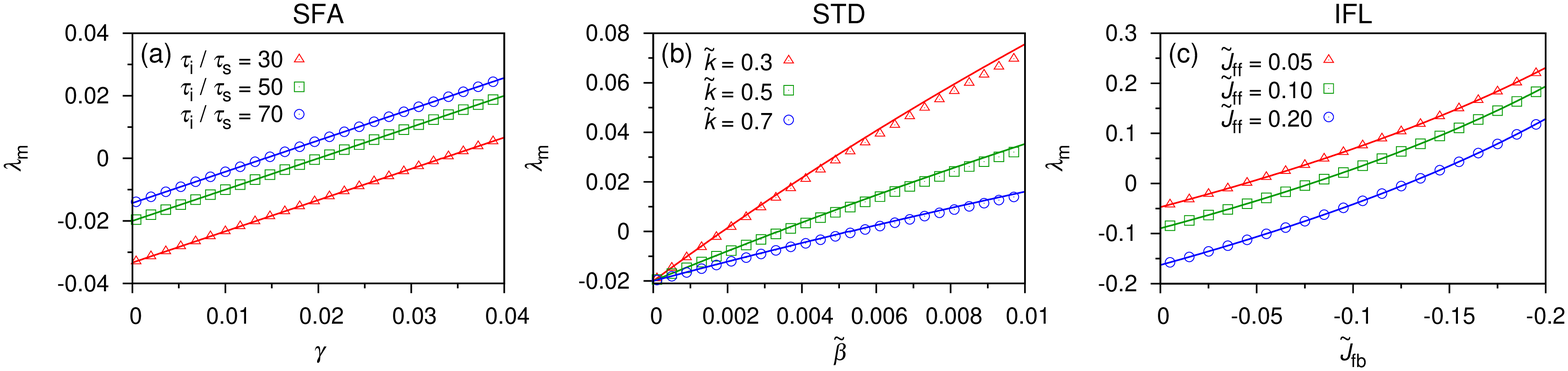} 
\par\end{centering}

\protect\protect\caption{\label{fig:lambda} (color online) The exponential rates of small
displacements of the $u$-profile from the $p$-profile, $\lambda$
for (a) SFA, (b) STD and (c) IFL. Symbols: simulations with various
combinations of parameters. Curves: prediction by Eq. (\ref{eq:propa_stability}).
Parameters: (a) $\tilde{k}=0.3$, (b) $\tau_{{\rm d}}=50\tau_{{\rm s}}$
and (c) $\tilde{k}=0.3$ and $\tau_{2}=\tau_{1}$. }
\end{figure*}

We have tested the prediction of Eq. (\ref{eq:propa_stability}) with
the three example models. In Fig. \ref{fig:lambda} there are simulation
results (symbols) plotted with the corresponding predictions (curves),
Eq. (\ref{eq:lambda_m}). In simulations the $p$-profile was intentionally
displaced by a tiny displacement from the $u$-profile after the system
has reached a stationary state. By monitoring the evolution of the
displacement, $\lambda$ can be measured. They agree with the prediction
very well. We can see that for small $\gamma$, $\tilde{\beta}$ and
$-\tilde{J}_{{\rm fb}}$, the displacement will decay to zero eventually.
But if these parameters are large enough, the tiny initial displacement
will diverge. This divergence of the displacement will eventually
lead to spontaneous motion. The results for SFA agree with those reported
by Mi \textit{et al.} \cite{Mi2014}, in which the system is able
to support spontaneously moving network state only when $\gamma>\tau_{s}/\tau_{i}$.


\begin{figure*}[t]
\begin{centering}
\includegraphics[width=1\textwidth]{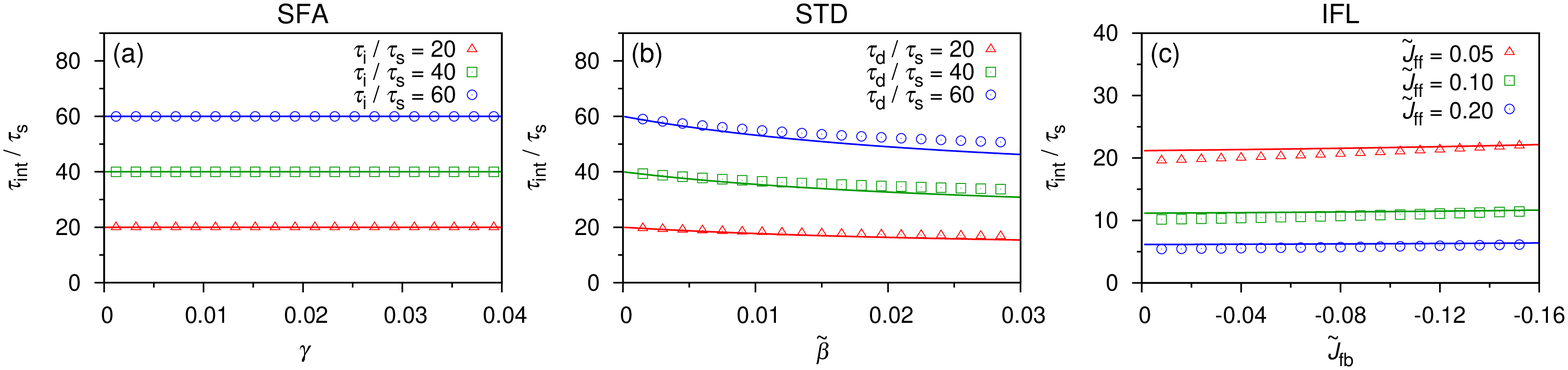} 
\par\end{centering}

\protect\protect\caption{\label{fig:tau_int} (color online) Comparison of the intrinsic time
scale measured with a moving stimulus probe (symbols) and theoretically
predicted (lines) for (a) SFA, (b) STD and (c) IFL. Parameters: (a)
-- (c) $\tilde{k}=0.3$ and $\hat{A}=0.25$. }
\end{figure*}

\section{Extrinsic Behavior}

In the presence of a weak and slow external stimulus, we consider
\begin{align}
u\left(x,t\right) & =u_{0}\left(x-v_{I}t\right),\\
p\left(x,t\right) & =p_{0}\left(x-v_{I}t\right)+\varepsilon_{0}\frac{dp_{0}(x-v_{I}t)}{dx},\\
I^{{\rm ext}}\left(x,t\right) & =\frac{\max_{x}u(x,t)}{\tau_{{\rm stim}}}\exp\left(-\frac{\left|x-v_{I}t+s\right|^{2}}{4a^{2}}\right).
\end{align}
Here $\tau_{{\rm stim}}$ is referred to as the stimulus time, representing
the time scale for the stimulus to produce significant response from
the exposed profile. $s$ is the displacement of the bump relative
to the stimulus. Substituting these assumptions into Eqs. (\ref{eq:dudt})
and (\ref{eq:dpdt}), we find that at the steady state of the weak
and slow stimulus limit, the separation $\varepsilon_{0}$ of the
exposed and inhibitory profiles is given by $\varepsilon_{0}=v_{I}\tau_{{\rm int}}$
to the lowest order, as derived in Appendix \ref{sec:tracking}. Since both $v_{I}$ and $\varepsilon_{0}$
can be measured in simulations, this provides a way to test the validity
of the theory. Indeed, simulations show that $\varepsilon_{0}$ is
linearly proportional to $v_{I}$, so that the slope can be compared
with the theoretical predictions of $\tau_{{\rm int}}$ by Eq. (\ref{eq:intrinsic_time}).
Results shown in Fig. \ref{fig:tau_int} for SFA, STD and IFL indicate
excellent agreement with theoretical predictions.

We further note that in Fig. \ref{fig:tau_int}, the values of $\tau_{{\rm int}}$
have been obtained for low values of $\gamma$, $\tilde{\beta}$ and
$-\tilde{J}_{{\rm fb}}$ where the bumps are intrinsically static.
A difference between the moving and static phases is that $\tau_{{\rm int}}$
can be deduced in the former via Eq. (\ref{eq:intrinsic_time}) whereas
the deduction is not possible in the latter since $v_{{\rm nat}}=0$.
Hence Fig. \ref{fig:tau_int} illustrates the close relation between
$\tau_{{\rm int}}$ measured extrinsically and intrinsically, and
that intrinsically inaccessible quantities can be obtained from extrinsic
measurements.

More relevant to the anticipatory phenomenon, we are interested in
the displacement $s$ and the anticipatory time $\tau_{{\rm ant}}$
of the exposed profile relative to the stimulus profile, given by

\begin{equation}
\tau_{{\rm ant}}\equiv\frac{s}{v_{I}}=\tau_{{\rm stim}}\tau_{{\rm int}}\lambda.\label{eq:tau_ant}
\end{equation}
The derivation can be found in Appendix \ref{sec:tracking}. Hence $\tau_{{\rm int}}$ and
$\lambda$ have the same sign. In the static phase, $\lambda<0$ implies
that the tracking is delayed with $\tau_{{\rm ant}}<0$, whereas in
the moving phase, $\lambda>0$ implies that the tracking is anticipatory
with $\tau_{{\rm ant}}>0$. At the phase boundary, $\lambda=0$ and
the system is in the ready-to-go state; here $\tau_{{\rm ant}}=0$
and the tracking is perfect.

Note that Eq. (\ref{eq:tau_ant}) is a manifestation of FRR, since
it relates the instability parameter $\lambda$, as an intrinsic property,
to the anticipatory time $\tau_{{\rm ant}}$, as an extrinsic property.
To see how this relation is consistent with 
traditional fluctuation-response relations, 
one should note that $\tau_{\rm ant}^{-1}$ describes 
the rate of response of the system to moving stimuli, 
and $\lambda^{-1}$ is proportional to fluctuations 
in both static and moving phases, 
as derived in Appendix \ref{sec:noise}.

For the example of the neural field with SFA in Fig. \ref{fig:moving_bump}(a),
the lag of the inhibitory profile $\tilde{p}$ drives the exposed
profile $\tilde{u}$ to move in the direction with smaller $\tilde{p}$
(pointed by the arrow), as $\tilde{p}$ inhibits $\tilde{u}$.

In the absence of SFA, the bell-shaped attractor state of $\tilde{u}$
centered at $z(t)$ (shown in Fig. \ref{fig:moving_bump}(b) as the
green dashed line) lags behind a continuously moving stimulus $z_{I}\left(t\right)$
(shown as the blue dotted line). In the inset of Fig.~\ref{fig:moving_bump}(b),
the lag of the network response develops after the stimulus starts
to move and becomes steady after a while. In contrast, when SFA is
sufficiently strong, the bump can track the stimulus at an advanced
position (red solid curve in Fig. \ref{fig:moving_bump}(b)). In this
case, this tracking process anticipates the continuously moving stimulus.
This behavior for SFA with various $\gamma$ and $v_{I}$ is summarized
in Fig. \ref{fig:moving_bump}(c).

Furthermore, the anticipation time is effectively constant in a considerable
range of the stimulus speed. There is an obvious advantage for the
brain to compensate delays with a constant leading time independent
of the stimulus speed. To put the speed independence of $\tau_{{\rm ant}}$
in a perspective, we note that $\varepsilon_{0}=v_{I}\tau_{{\rm int}}$,
implying that $\tau_{{\rm ant}}=\lambda\tau_{{\rm stim}}\varepsilon_{0}/v_{I}$.
This shows that while the stimulus speed increases, the lag of the
inhibitory profile behind the exposed profile also increases, providing
an increasing driving force for the bump such that the anticipatory
time remains constant.

This is confirmed when the SFA strength $\gamma$ is strong enough.
As shown in Fig.~\ref{fig:moving_bump}(c) for $\gamma=0.1$, there
is a velocity range such that the displacement of the center of mass
relative to the stimulus, $z(t)-z_{I}(t)$, is directly proportional
to the stimulus velocity. Thus the anticipation time $\tau_{{\rm ant}}\equiv(z-z_{I})/v_{I}$,
given by the slope of the curve, is effectively constant. In Fig.~\ref{fig:moving_bump}(d),
the anticipatory time is roughly 0.3$\tau_{{\rm i}}$ ($\tau_{{\rm i}}$
is the time constant of SFA) for a range of stimulus velocity, and
has a remarkable fit with data from rodent experiments \cite{Goodridge2000}.
This behavior can also be observed in neural field models with STD
\cite{Fung2013}.


\begin{figure*}[t]
\begin{centering}
\includegraphics[width=1\textwidth]{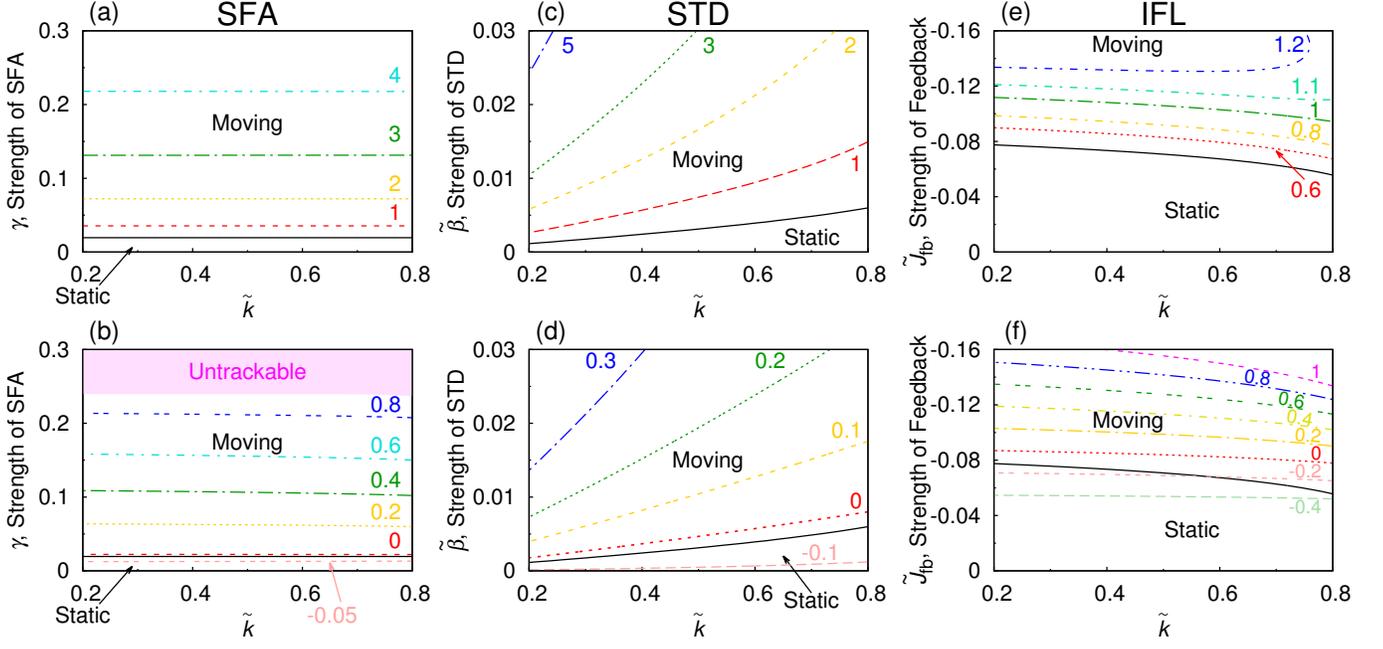} 
\par\end{centering}

\protect\protect\caption{\label{fig:spd_n_tauant} (color online) (a) Contours of intrinsic
speed in the phase diagram of a neural field model with SFA. (b) Contours
of anticipation time of a neural field model with SFA. (c) Same as
(a), but for STD. (d) Same as (b), but for STD. (e) Same as (a), but
for IFL. (f) Same as (b), but for IFL. Color curves: contours of intrinsic
speed ((a), (c) \& (e)), anticipatory time ((b), (d) \& (f)). Number
labels: values of the corresponding contour, in units of (a) $a/\tau_{{\rm i}}$,
(b) $\tau_{{\rm i}}$, (c) $a/\tau_{{\rm d}}$, (d) $\tau_{{\rm d}}$
(e) $\tau_{2}/\tilde{J}_{{\rm ff}}$ and (f) $a/(\tau_{2}/\tilde{J}_{{\rm ff}})$.
Black curves: phase boundaries separating the static, moving, and
silent phases. Parameters: (a) $\tau_{{\rm i}}=50\tau_{{\rm s}}$.
(b) $\hat{A}=0.25$, $v_{I}=0.002a/\tau_{{\rm s}}$ and $\tau_{{\rm i}}=50\tau_{{\rm s}}$.
(c) $\tau_{{\rm d}}$~(time constant of STD) $=50\tau_{{\rm s}}$.
(d) $\hat{A}=0.25$, $v_{I}=0.002a/\tau_{{\rm s}}$ and $\tau_{{\rm d}}=50\tau_{{\rm s}}$.
(e) $\tilde{J}_{{\rm ff}}=0.1$ and $\tau_{1}$~(time constant of
the primary layer) $=\tau_{2}$~(time constant of the hidden layer)
$=\tau_{{\rm s}}$, (f) $\tilde{J}_{{\rm ff}}=0.1$, $\hat{A}=0.1$,
$v_{I}=0.002a/\tau_{{\rm s}}$, $\tau_{1}=\tau_{2}$. In the shaded
area of (b), $\hat{A}$ is too small to stabilize the system. One
should note that metastatic phase reported in \cite{Fung2012} for
STD are omitted in the current study, as the major concern in the
paper is the relation between translational intrinsic behavior and
translational extrinsic behavior.}
\end{figure*}

The interdependency of anticipatory tracking dynamics and intrinsic
dynamics in the framework of FRR is further illustrated by the relation
between the anticipatory time and the intrinsic speed of spontaneous
motions. Near the boundary of the moving phase, it is derived in Appendix
\ref{sec:tracking} that 
\begin{equation}
\tau_{{\rm ant}}=K\tau_{{\rm stim}}\tau_{{\rm int}}\left(v_{{\rm nat}}^{2}-v_{I}^{2}\right)+\tau_{{\rm con}},\label{eq:tau_ant_general}
\end{equation}
or the quadratic relation in the limit of weak and slow stimulus
\begin{equation}
\tau_{{\rm ant}}=K\tau_{{\rm stim}}\tau_{{\rm int}}v_{{\rm nat}}^{2},
\end{equation}
where $K$ and $\tau_{{\rm con}}$ are constants defined in Appendix 
\ref{sec:tracking}. Since all parameters besides $v_{{\rm nat}}^{2}$ and 
$v_{I}^{2}$ (taken to approach 0) are mostly slowly changing functions of 
system parameters, the contours of $v_{{\rm nat}}$ and $\tau_{{\rm ant}}$ 
in the parameter space have a one-to-one correspondence. The case for SFA 
is illustrated in Fig. \ref{fig:spd_n_tauant}(a) and (b).


\begin{figure*}[t]
\begin{centering}
\includegraphics[width=0.9\textwidth]{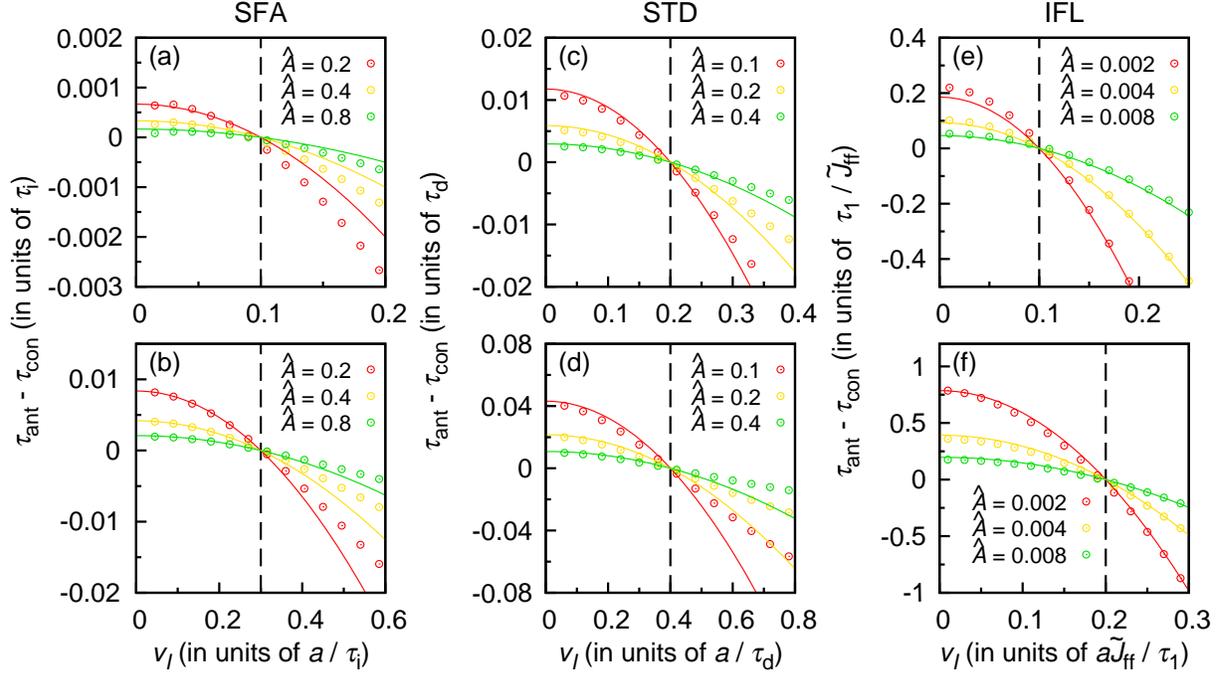} 
\par\end{centering}

\protect\protect\caption{\label{fig:s_of_v} (color online) Anticipatory time versus the speed
of the stimulus $v$. Black dashed lines: intrinsic speed of the corresponding
set of parameters. Parameters: (a) $\tilde{k}=0.3$, $\gamma=0.0202$,
$\tau_{i}=50\tau_{s}$ and $\hat{A}$ is labeled along with curves.
(b) $\tilde{k}=0.3$, $\gamma=0.0217$,and $\tau_{i}=50\tau_{s}$.
(c) $\tilde{k}=0.3$, $\tilde{\beta}=0.00198$ and $\tau_{{\rm d}}=50\tau_{s}$.
(d) $\tilde{k}=0.3$, $\tilde{\beta}=0.00231$ and $\tau_{{\rm d}}=50\tau_{s}$.
(e) $\tilde{k}=0.6$, $\tilde{J}_{{\rm fb}}=-0.0698$ and $\tau_{2}=\tau_{1}$.
(f) $\tilde{k}=0.6$, $\tilde{J}_{{\rm fb}}=-0.0705$ and $\tau_{2}=\tau_{1}$. }
\end{figure*}

Since these phenomena depend on the underlying symmetry of the system
and its response to weak stimuli, they are expected to be observed
in networks with the same symmetry as SFA networks. The correspondence
between intrinsic motion and anticipation has been described in the
specific case of STD networks \cite{Fung2013}. Comparable contour
plots to Fig. \ref{fig:spd_n_tauant}(a) and (b) for STD are shown
in \ref{fig:spd_n_tauant}(c) and (d), respectively. Similar phenomena
can be found in Fig. \ref{fig:spd_n_tauant}(e) and (f) for IFL, except
that the contours in Fig.~\ref{fig:spd_n_tauant} are distorted in
the proximity of the repulsive phase (Repulsive phase can be observed
if $(-\tilde{J}_{{\rm fb}})\gg\tilde{J}_{{\rm ff}}$, see Appendix 
\ref{sec:IFL} for more details). A minor discrepancy is that the contour 
for zero anticipatory
time does not coincide perfectly with the phase boundary separating
the moving and static phases. This is due to deviations from the weak
input limit, since a finite input amplitude is necessary to prevent
the network state from becoming ``untrackable''. For SFA, the untrackable
region is shaded in Fig. \ref{fig:spd_n_tauant}(b). For IFL, the
untrackable region is located immediately beyond the upper right corner
of Fig. \ref{fig:spd_n_tauant}(f).

\section{Natural Tracking}

For non-vanishing stimulus velocities in the moving phase, Eq.~(\ref{eq:tau_ant_general})
predicts another interesting phenomenon linking tracking dynamics
and intrinsic dynamics. When the stimulus is moving at the natural
speed, i.e. $v_{I}=v_{{\rm nat}}$, the anticipatory time becomes
independent of the strength of the external input which determines
$\tau_{{\rm stim}}$, and the anticipation time curves are confluent
at the value $\tau_{{\rm ant}}=\tau_{{\rm con}}$. This phenomenon
for a particular neural field model with STD has been reported in
\cite{Fung2013}; here we show that it is generic in an entire family
of neural fields.

The physical picture of this confluent behavior is that the stimulus
plays two roles in driving the moving bump. First, it is used to drive
the bump at the stimulus speed, if it is different from the intrinsic
speed. Second, it is used to distort the shape of the bump. In the
second role, the distortion is proportional to both the strength of
the stimulus and the bump-stimulus displacement, $z(t)-z_{0}(t)$.
Hence when the stimulus speed is the same as the intrinsic speed,
the stimulus is primarily used to distort the bump shape. At the steady
state, the bump-stimulus displacement is determined by the distortion
per unit stimulus strength, which becomes independent of stimulus
strength.

Since this phenomenon is based on a generic mechanism, it can be observed
in all neural field models considered in this paper. Fig. \ref{fig:s_of_v}
shows the simulation results in neural field models with SFA, STD
and IFL. Fig. \ref{fig:s_of_v}(a) shows the displacements in the
SFA neural field model with the intrinsic speed $v_{{\rm nat}}=0.1a/\tau_{{\rm i}}$,
where $\tau_{{\rm i}}$ is the SFA time scale. $\tau_{{\rm ant}}$-$v_{I}$
curves corresponding to different stimulus amplitudes intersect at
$\tau_{{\rm i}}v_{{\rm nat}}/a=0.1$. Similar behaviors are shown
in Fig. \ref{fig:s_of_v}(b) for $v_{{\rm nat}}=0.3a/\tau_{{\rm i}}$,
in Fig. \ref{fig:s_of_v}(c) and (d) for STD, and in Fig. \ref{fig:s_of_v}(e)
and (f) for IFL. Remarkably, the confluent behavior remains valid
even when the curves deviate from the parabolic shape predicted by
Eq. (\ref{eq:tau_ant_general}).

\section{Noise Response}

To further illustrate FRR, we consider the correlation between fluctuations
due to noise in the absence of external input and the anticipatory
time reacting to a weak and slow moving stimulus. This can be done
by replacing $I^{{\rm ext}}$ in Eq. (\ref{eq:dudt}) with displacement
noise $\xi\left(x,t\right)\equiv\eta\left(t\right)\partial u_{0}/\partial x$,
where $\left\langle \eta\left(t\right)\right\rangle =0$ and $\left\langle \eta\left(t\right)\eta\left(t^{\prime}\right)\right\rangle =2T\delta\left(t-t^{\prime}\right)$.
Analysis in Appendix \ref{sec:noise} shows that for weak and slow stimuli,

\begin{equation}
\frac{\left\langle \delta\varepsilon_{0}^{2}\right\rangle }{T}=\begin{dcases}
-\frac{\tau_{{\rm stim}}\tau_{{\rm int}}}{\tau_{{\rm ant}}-\tau_{{\rm con}}},
&\mbox{for static phase,}\\
\frac{\tau_{{\rm stim}}\tau_{{\rm int}}}{2(\tau_{{\rm ant}}-\tau_{{\rm con}})},
&\mbox{for moving phase.}
\end{dcases}
\label{eq:response}
\end{equation}
Here, $\langle\delta\varepsilon_{0}^{2}\rangle$ represents the fluctuations
of the lag of the inhibitory profile $p(x,t)$ behind the exposed
profile $u(x,t)$ in response to the displacement noise.


\begin{figure*}
\begin{centering}
\includegraphics[width=1\textwidth]{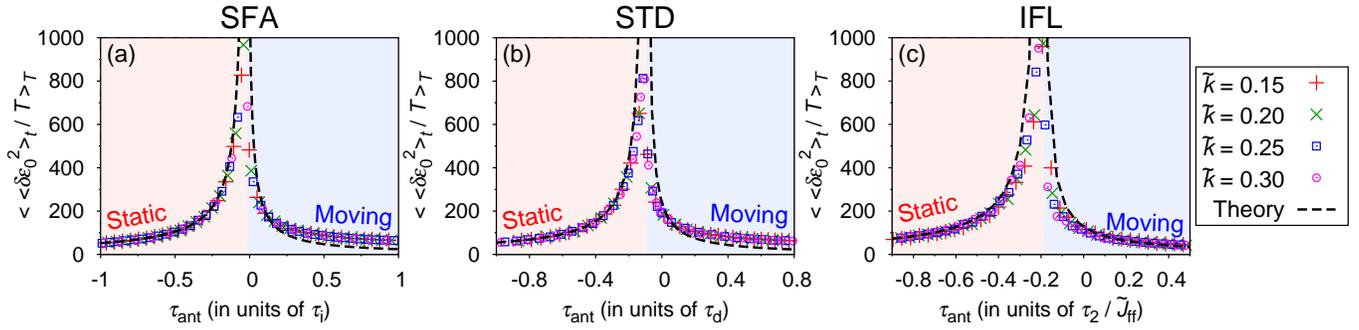} 
\par\end{centering}

\protect\protect\caption{\label{fig:response} (color online) Intrinsic noise response of the
system, $\left\langle \left\langle \delta\varepsilon_{0}^{2}\right\rangle _{t}/T\right\rangle _{T}$,
versus anticipation time, $\tau_{{\rm ant}}$. Parameters: (a) $\hat{A}=0.02$
and $\tau_{i}=50\tau_{s}$. (c) $\hat{A}=0.02$ and $\tau_{d}=50\tau_{s}$.
(b) $\hat{A}=0.02$, $\tilde{J}_{{\rm ff}}=0.1$ and $\tau_{2}=\tau_{1}$. }
\end{figure*}

The behavior predicted by Eq.~(\ref{eq:response}) can be seen from simulations. 
The numerical procedure is explained 
in Appendix \ref{sec:numeric}.
In Fig. \ref{fig:response}, there are two branches in
each sub-figure. The branches for $\tau_{{\rm ant}}>\tau_{{\rm con}}$
and $\tau_{{\rm ant}}<\tau_{{\rm con}}$ correspond to the moving
and static phases respectively. Remarkably, data points with different
network parameters collapse onto common curves. The fluctuations are
divergent at the confluence point predicted by Eq.~(\ref{eq:tau_ant_general}).
The regimes of $\tau_{{\rm ant}}>0$ and $\tau_{{\rm ant}}<0$, corresponding
to anticipatory and delayed tracking respectively, effectively coincide
with the two branches in the limit of weak stimuli, since at the confluence
point the instability eigenvalue $\lambda=(\tau_{{\rm ant}}-\tau_{{\rm con}})/(\tau_{{\rm stim}}\tau_{{\rm int}})$
approaches 0 in that limit.

\section{Conclusion}

Many intriguing dynamical behaviors of physical systems can be understood
from the relationship between the fluctuation properties of a system
near equilibrium and its response to external driving fields, namely,
the FRR \cite{Einstein1905,vonSmoluchowski1906,Nyquist1928,Huang1987}.
Here, we show that the same idea is applicable to understanding the
dynamics of neural fields. In particular, we have found a fluctuation-response
relation for neural fields processing dynamical information. Traditionally,
theoretical techniques based on equilibrium concepts have been well
developed in analyzing neural fields processing static information.
On the other hand, neural fields responding to external dynamical
information are driven to near-equilibrium states, and FRRs are suitable
tools to describe their behaviors.

There have been previous analyses on neural fields with slow, localized
inhibitory feedbacks. Moving phases and anticipatory tracking have
been studied in neural fields with STD \cite{York2009,Fung2012,Fung2013,Fung2015},
SFA \cite{Mi2014,Fung2015} and IFL \cite{Ben-Yishai1997,Zhang2012}.
However, results of the boundary between the static and moving phases,
the intrinsic speed or the tracking delay were specific to the particular
models, concealing their common underlying physical principles.

The unification of these various manifestations were provided by the
FRR considered in this paper. We pointed out that they have a common
structure consisting of an exposed variable ($u$) coupled to external
stimuli and an inhibitory variable ($p$) hidden from stimuli. Irrespective
of the explicit form of the dynamical equations, the FRR is generically
based on (i) the existence of a non-zero solution, and (ii) this solution
is translationally invariant and (iii) possesses inversion symmetry.
Consequently, FRR is able to relate (i) the positional stability of
the activity states, to (ii) their lagging/leading position relative
to external stimuli during tracking, and to (iii) fluctuations due
to thermal noises.

Particularly relevant to the processing of motional information, FRR
predicts that the regimes of anticipatory and delayed tracking effectively
coincide with the regimes of moving and static phases respectively,
and that the anticipation time becomes independent of stimulus speed
for slow and weak stimuli, and independent of stimulus amplitude when
the stimulus moves at the intrinsic speed.

This brings FRR into contact with experimental observations of how
neural systems cope with time delays in the transmission and processing
of signals, which are ubiquitous in neural systems. To compensate
for delays, neural systems need to anticipate moving stimuli, which
has been observed in HD cells of rodents \cite{Goodridge2000}. FRR
provides the condition for the anticipatory behavior. Furthermore,
we predict that the anticipatory time is independent of the stimulus
speed, offering the advantage of a fixed time for the system to respond.

FRR also provides a means to measure quantities that are normally
inaccessible in certain regimes. For example, the intrinsic time in
the static phase is intrinsically unmeasurable since there is no separation
between the exposed and inhibitory profiles in that phase. Our analysis
shows that the intrinsic time is identical to the local time lapse
between the exposed and inhibitory profiles due to moving stimuli,
thus providing an extrinsic instrument to measure the intrinsic time.


Since FRR is successful in unifying the behaviors of neural fields
with slow inhibitory feedback mechanisms such as STD, SFA, IFL and
other neural fields of the family, it can be extended to study the
relation between fluctuations and responses in other modes of encoding
information, such as amplitude fluctuations and amplitude responses.
It is expected to be an important element in understanding the processing
of dynamical information in the brain. It can also be applied to other
natural or artificial dynamical systems in which motional information
needs to be processed in real time, and FRR provides a powerful tool
to analyze the dynamical properties of these systems.
\begin{acknowledgments}
This work is supported by the Research Grants Council of Hong Kong
(grant numbers 604512, 605813 and N\texttt{\char`_}HKUST606/12), the National Foundation
of Natural Science of China (No. 31221003, No. 31261160495) and the
973 program (2014CB846101) of Ministry of Science and Technology of
China. 
\end{acknowledgments}

\appendix

\renewcommand{\thefigure}{A.\arabic{figure}}
\setcounter{figure}{0}

\section{Intrinsic Behaviors of Inhibitory Feedback Loops\label{sec:IFL}}

This is one of the three examples mentioned in the main text. For
the other two examples, a detailed study on CANNs with STD can be
found in \cite{Fung2012}, and the intrinsic behavior of CANNs with
SFA is similar. In this section, the intrinsic behaviors of a bump-shaped
profile in a two-layered network with an inhibitory feedback loop
(IFL) are summarized.

\begin{figure}
\begin{centering}
\includegraphics[width=\columnwidth]{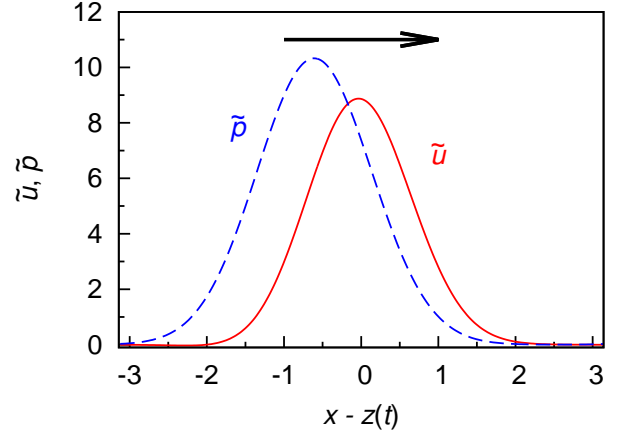}
\par\end{centering}

\protect\caption{\label{fig:moving_tl} (color online) A snapshot of the network state of a two-layered
network in its moving phase. $\tilde{u}$ and $\tilde{p}$
are the rescaled neuronal current profile of the first and second
layers respectively. The profiles are moving in the direction of the
arrow at the top. Parameters: $\tilde{k}=0.5$, $\tilde{J}_{{\rm ff}}=0.1$,
$\tilde{J}_{{\rm fb}}=-0.1$ and $\tau_{2}=\tau_{1}=1$.}
\end{figure}

If the negative feedback strength ($\tilde{J}_{{\rm fb}}$) is strong
enough, the bump in the second layer that provides a negative feedback
to the first layer can destabilize the bump in the first layer. At
the steady state, the misalignment between two profiles becomes a
constant. As shown in Fig. \ref{fig:moving_tl}, the two misaligned
bumps move spontaneously. Since the neurons in the first layer receive
negative feedbacks and neurons in the second layer receives positive
feedforwards, the magnitude of $\tilde{p}$-profile is larger
than $\tilde{u}$-profile.

\begin{figure*}
\begin{centering}
\includegraphics[width=\textwidth]{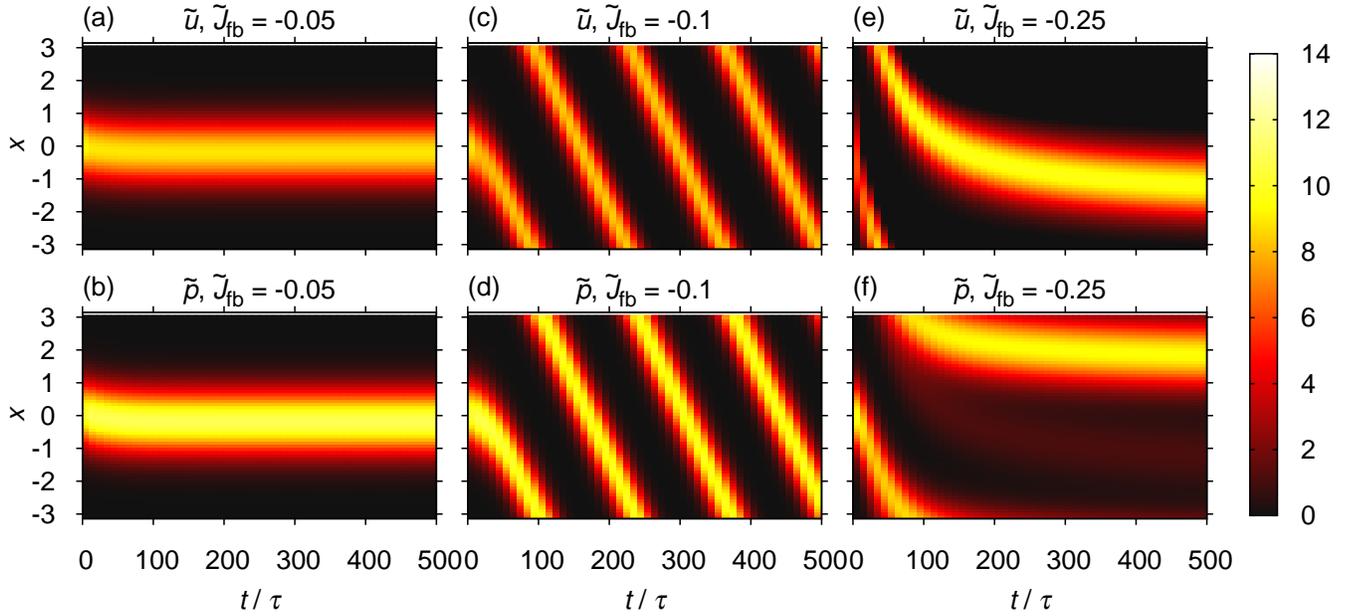}
\par\end{centering}

\protect\caption{\label{fig:states_tl} (color online) Typical examples of network behaviors for various
phases: static phase, moving phase and repulsive phase. (a) and (b):
static phase. (c) and (d): moving phase. (e) and (f): repulsive phase.
Parameters: $\tilde{k}=0.5$, $\tilde{J}_{{\rm ff}}=0.1$ and $\tau_{2}=\tau_{1}=1$.}
\end{figure*}

The intrinsic behavior supported by the system is determined by the
choice of parameters. Figure \ref{fig:states_tl} shows the typical
cases of the static phase, the moving phase and the repulsive phase.
In simulations, the initi al conditions of $\tilde{u}$ and $\tilde{p}$
are misaligned so that the environment of $\tilde{u}$ is not
symmetric about its center. If the magnitude of $\tilde{J}_{{\rm fb}}$
is not strong enough, the bump will relax to a static state, see Fig.
\ref{fig:states_tl} (a) and (b). For a sufficiently strong $\tilde{J}_{{\rm fb}}$,
the bump can move spontaneously as in Fig. \ref{fig:moving_tl} and
Fig. \ref{fig:states_tl} (c) and (d). This is the moving phase. In
this phase, the $\tilde{p}$-profile repels the $\tilde{u}$-profile.
However, at the same time, the $\tilde{u}$-profile attracts the
$\tilde{p}$-profile. So, at the equilibrium state, the misalignment
between two profiles becomes steady.

If $\tilde{J}_{{\rm fb}}$ is too strong, the spontaneous motion will
terminate. In this case, initially, the $\tilde{p}$-profile repels
the $\tilde{u}$-profile and the $\tilde{u}$-profile attracts
the $\tilde{p}$-profile. However, in the repulsive phase, the
repulsion is so strong that the attraction can no longer balance the
repulsive force. As a result, the two profiles move apart out of the
interactive range of each other, as shown in Fig. \ref{fig:states_tl}
(e) and (f). The spontaneous motion cannot sustain at the steady state.
In general, together with the trivial solution, there are four phases
in two-layer CANNs, under the current setting. The phase diagram for
these four phases is shown in the main paper.

The slowness of the inhibitory feedback, and hence the existence of
the moving phase, arises from the weak coupling between the exposed
and inhibitory layers. To see this, we consider the moving bump solution
\begin{align}
u(x,t)&=u^0\exp\left[-\frac{(x-vt)^{2}}{4a^{2}}\right],\quad\mathrm{and}\\
p(x,t)&=p^0\exp\left[-\frac{(x-vt+s)^{2}}{4a^{2}}\right].
\end{align}
Substituting into Eq. (\ref{eq:dudt}), multiplying both sides by $\exp[-(x-vt)^{2}/(4a^{2})]/\sqrt{2\pi a^{2}}$
and integrating,
\begin{equation}
\tilde{u}^0=\frac{\left.\tilde{u}^0\right.^{2}}{\sqrt{2}B_{u}}+\tilde{J}_{\mathrm{fb}}\frac{\left.\tilde{p}^0\right.^{2}}{\sqrt{2}B_{p}}e^{-\frac{s^{2}}{8a^{2}}},
\end{equation}
where $B_{u}=1+\tilde{k}\left.\tilde{u}^0\right.^{2}/8$ and $B_{p}=1+\tilde{k}\left.\tilde{p}^0\right.^{2}/8$.

Substituting into Eq. (\ref{eq:dudt}), multiplying both sides by $[(x-vt)/a]\exp[-(x-vt)^{2}/(4a^{2})]/\sqrt{2\pi a^{2}}$
and integrating,
\begin{equation}
\frac{v\tau_{1}}{2a}\tilde{p}^0=-\tilde{J}_{{\rm fb}}\frac{\left.\tilde{p}^0\right.^{2}}{\sqrt{2}B_{p}}\frac{s}{2a}e^{-\frac{s^{2}}{8a^{2}}}.
\end{equation}
Consider the condition for the moving phase boundary with both $v$
and $s$ approaching 0 at a finite ratio. The above equations imply
that
\begin{equation}
\frac{v\tau_{1}}{s}=-\frac{\tilde{J}_{{\rm fb}}\frac{\left.\tilde{p}^0\right.^{2}}{\sqrt{2}B_{p}}}{\frac{\left.\tilde{u}^2\right.^{2}}{\sqrt{2}B_{u}}+\tilde{J}_{\mathrm{fb}}\frac{\left.\tilde{p}^0\right.^{2}}{\sqrt{2}B_{p}}}\sim-\frac{\tilde{J}_{{\rm fb}}}{1+\tilde{J}_{{\rm fb}}}.
\end{equation}
Similarly, by considering the dynamics of the second layer, we have
\begin{equation}
\frac{v\tau_{2}}{s}\sim\frac{\tilde{J}_{{\rm ff}}}{1+\tilde{J}_{{\rm ff}}}.
\end{equation}

Hence weak interlayer couplings, $\left|\tilde{J}_{{\rm fb}}\right|\ll1$
or $\tilde{J}_{{\rm ff}}\ll1$ play the same role as the ratio $\tau_{{\rm s}}/\tau_{{\rm d}}$
in STD \cite{Fung2012}.

\section{Intrinsic Behavior of Profile Separation\label{sec:separation} }

We consider perturbations that cause the exposed and inhibitory profiles
to separate. These distortions have odd parity. To keep the discussions
general, we further assume that distortion modes with even parity
also contribute to the perturbations. As we shall see, the coupling
of these even parity modes with the odd parity modes play a role in
determining the intrinsic and extrinsic behaviors in the moving phase.
Hence we consider perturbations of the form 
\begin{equation}
\delta u\left(x\right)=c_{0}\frac{\partial u_{0}}{\partial x}+c_{1}u_{1}\left(x\right)\text{{\rm , }}\delta p\left(x\right)=\varepsilon_{0}\frac{\partial p_{0}}{\partial x}+\varepsilon_{1}p_{1}\left(x\right).\label{eq:deltau_deltap_intrinsic_beh}
\end{equation}
$c_{0}$ and $\varepsilon_{0}$ are considered to be the displacement
of the exposed and inhibitory profiles respectively (in the direction
opposite to their signs). $u_{1}$ and $p_{1}$ are the most significant
even parity distortion modes. They are substituted into the dynamical
equations (\ref{eq:depsilondcdt_a}) and (\ref{eq:depsilondcdt_b}). Multiplying both sides of Eq. (\ref{eq:depsilondcdt_a}) by $g_{u}^{0}$
and integrating, 
\begin{align}
{\rm LHS}&=\frac{\partial c_{0}}{\partial t}\int dxg_{u}^{0}\left(x\right)\frac{\partial u_{0}}{\partial x}+\frac{\partial c_{1}}{\partial t}\int dxg_{u}^{0}\left(x\right)u_{1}\left(x\right) \nonumber \\
&=\frac{\partial c_{0}}{\partial t}I_{u},
\label{eq:LHS_ddeltaudt}
\end{align}
where, for $i=u$, $p$, 
\begin{equation}
I_{i}=\int dxg_{i}^{0}\left(x\right)\frac{\partial u_{i}^{0}}{\partial x},
\end{equation}

Note that the second term in Eq. (\ref{eq:LHS_ddeltaudt}) vanishes
since $g_{u}^{0}$ and $u_{1}$ have opposite parity. On the right
hand side, 
\begin{align}
{\rm RHS}_{1}&=c_{0}\int dxg_{u}^{0}\left(x\right)\int dx^{\prime}\frac{\partial F_{u}\left(x\right)}{\partial u\left(x^{\prime}\right)}\frac{\partial u_{0}\left(x^{\prime}\right)}{\partial x^{\prime}}\nonumber\\
&+c_{1}\int dxg_{u}^{0}\left(x\right)\int dx^{\prime}\frac{\partial F_{u}\left(x\right)}{\partial u\left(x^{\prime}\right)}u_{1}\left(x^{\prime}\right).
\end{align}
The second term vanishes due to odd parity. Hence 
\begin{equation}
{\rm RHS}_{1}=c_{0}\int dxg_{u}^{0}\left(x\right)\int dx^{\prime}\frac{\partial F_{u}\left(x\right)}{\partial u\left(x^{\prime}\right)}\frac{\partial u_{0}\left(x^{\prime}\right)}{\partial x^{\prime}}=c_{0}Q_{uu}
\end{equation}
Similarly, the second term on the right hand side becomes 
\begin{equation}
{\rm RHS}_{2}=\varepsilon_{0}\int dxg_{u}^{0}\left(x\right)\int dx^{\prime}\frac{\partial F_{u}\left(x\right)}{\partial p\left(x^{\prime}\right)}\frac{\partial p_{0}\left(x^{\prime}\right)}{\partial x^{\prime}}=\varepsilon_{0}Q_{up}.
\end{equation}
Hence we obtain 
\begin{equation}
I_{u}\frac{\partial c_{0}}{\partial t}=Q_{uu}c_{0}+Q_{up}\varepsilon_{0}.\label{eq:dc0dt_intrinsic_beh}
\end{equation}
Similarly, from Eq. (\ref{eq:depsilondcdt_b}), 
\begin{equation}
I_{p}\frac{\partial\varepsilon_{0}}{\partial t}=Q_{pu}c_{0}+Q_{pp}\varepsilon_{0}.\label{eq:de0dt_intrinsic_beh}
\end{equation}
Using the identities of translational invariance in Eqs. (\ref{eq:e1_speed}) and (\ref{eq:K_i}), 
\begin{equation}
\frac{\partial}{\partial t}\left(\begin{array}{c}
c_{0}\\
\varepsilon_{0}
\end{array}\right)=\left(\begin{array}{cc}
Q_{uu}/I_{u} & -Q_{uu}/I_{u}\\
-Q_{pp}/I_{p} & Q_{pp}/I_{p}
\end{array}\right)\left(\begin{array}{c}
c_{0}\\
\varepsilon_{0}
\end{array}\right).
\end{equation}
This implies
\begin{eqnarray}
\frac{\partial}{\partial t}\left(\varepsilon_{0}-c_{0}\right) & = & \left(\frac{Q_{uu}}{I_{u}}+\frac{Q_{pp}}{I_{p}}\right)\left(\varepsilon_{0}-c_{0}\right),\nonumber \\\label{eq:depsilondcdt_a}\\
\frac{\partial}{\partial t}\left(\frac{I_{u}}{Q_{uu}}c_{0}+\frac{I_{p}}{Q_{pp}}\varepsilon_{0}\right) & = & 0.\label{eq:depsilondcdt_b}
\end{eqnarray}
Eq. (\ref{eq:depsilondcdt_a}) describes the dynamics of the displacement
of the inhibitory profile relative to the exposed profile. The instability
eigenvalue in Eq. (\ref{eq:depsilondcdt_a}) is denoted as

\begin{equation}
\lambda\equiv\frac{Q_{uu}}{I_{u}}+\frac{Q_{pp}}{I_{p}}.\label{eq:eigenvalue}
\end{equation}

\section{Intrinsic Speed\label{sec:speed} }

When the bump becomes translationally unstable, it moves with an intrinsic
speed (or natural speed). To investigate the intrinsic speed, we need
to expand the dynamical equation beyond first order. In this case,
the translational variables become coupled with the next eigenfunction.
To keep the analysis trackable, we choose the coordinate with $c_{0}=0$.
Near the phase boundary of the static and moving phases, $v_{\mathrm{nat}}\sim\varepsilon_{0}$
and $c_{1}\sim\varepsilon_{1}\sim\varepsilon_{0}^{2}$, as will be
verified in this section. Hence to include third order terms, it is
sufficient to consider terms in the dynamical equations containing
$\varepsilon_{{\rm 0}}$, $c_{1}$, $\varepsilon_{1}$, $\varepsilon_{0}^{2}$,
$\varepsilon_{{\rm 0}}c_{1}$, $\varepsilon_{0}\varepsilon_{1}$,
$\varepsilon_{0}^{3}$, $v_{\mathrm{nat}}\varepsilon_{0}$, $v_{\mathrm{nat}}c_{1}$,
$v_{\mathrm{nat}}\varepsilon_{1}$. Substituting Eq. (\ref{eq:deltau_deltap_intrinsic_beh})
into the dynamical equation (\ref{eq:depsilondcdt_a}), expanding to third order for a bump
moving with natural speed $v_{\mathrm{nat}}$, multiplying both sides
of Eq. (\ref{eq:depsilondcdt_a}) by $g_{u}^{0}$ and integrating,
\begin{align}
-I_{u}v_{\mathrm{nat}}-M_{u}v_{\mathrm{nat}}c_{1}=~&Q_{up}\varepsilon_{{\rm 0}}+T_{upu}\varepsilon_{0}c_{1}+T_{upp}\varepsilon_{0}\varepsilon_{1}\nonumber\\&+\frac{Q_{uppp}}{6}\varepsilon_{0}^{3},\label{eq:c0_speed}
\end{align}
where, for $i$, $j$, $k$, $l=u$, $p$, 
\begin{align}
M_{i} & = \int dxg_{i}^{0}\frac{\partial u_{i}^{1}(x)}{\partial x},\\
T_{ijk} & = \int dxg_{i}^{0}\left(x\right)\int dx_{1}\int dx_{2}\nonumber \\
&\qquad\qquad\frac{\partial^{2}F_{i}\left(x\right)}{\partial u_{j}\left(x_{1}\right)\partial u_{k}\left(x_{2}\right)}\frac{\partial u_{j}^{0}\left(x_{1}\right)}{\partial x_{1}}u_{k}^{1}(x_{2}),\\
Q_{ijkl} & = \int dxg_{i}^{0}\left(x\right)\int dx_{1}\int dx_{2}\int dx_{3}\nonumber\\
&\qquad\qquad\frac{\partial^{3}F_{i}\left(x\right)}{\partial u_{j}\left(x_{1}\right)\partial u_{k}\left(x_{2}\right)\partial u_{l}\left(x_{3}\right)}\frac{\partial u_{j}^{0}\left(x_{1}\right)}{\partial x_{1}}\nonumber\\
&\qquad\qquad\qquad\frac{\partial u{}_{k}^{0}\left(x_{2}\right)}{\partial x_{2}}\frac{\partial u_{l}^{0}\left(x_{3}\right)}{\partial x_{3}}.
\end{align}
The left hand side of Eq. (\ref{eq:c0_speed}) arises from the time
rate of change of the neural activities at a location when the bump
passes by. These terms are proportional to the bump velocity and are
referred to as the wave terms. Substituting Eq. (\ref{eq:deltau_deltap_intrinsic_beh})
into the dynamical equation (\ref{eq:depsilondcdt_a}), multiplying both sides of Eq. (\ref{eq:depsilondcdt_a})
by $g_{u}^{1}$ and integrating, 
\begin{equation}
{\rm LHS}=\frac{\partial c_{1}}{\partial t}\int dxg_{u}^{1}\left(x\right)u_{1}\left(x\right)=\frac{\partial c_{1}}{\partial t}J_{u},
\end{equation}
where, for $i=u$, $p$, 
\begin{equation}
J_{i}=\int dxg_{i}^{1}\left(x\right)u_{i}^{1}\left(x\right),
\end{equation}
with $u_{i}^{1}(x)$ representing the functions $u_{1}(x)$ and $p_{1}(x)$
for $i=u$, $p$ respectively. On the right hand side,
\begin{align}
{\rm RHS} = &~  c_{1}\int dxg_{u}^{1}\left(x\right)\int dx^{\prime}\frac{\partial F_{u}\left(x\right)}{\partial u\left(x^{\prime}\right)}u_{1}\left(x^{\prime}\right) \nonumber \\ 
&+\varepsilon_{1}\int dxg_{u}^{1}\left(x\right)\int dx^{\prime}\frac{\partial F_{u}\left(x\right)}{\partial p\left(x^{\prime}\right)}p_{1}\left(x^{\prime}\right)\nonumber \\
 &   +\frac{\varepsilon_{{\rm 0}}^{2}}{2}\int dx_{1}\int dx_{2}\frac{\partial^{2}F_{u}(x)}{\partial p(x_{1})\partial p(x_{2})}\frac{\partial p_{0}(x_{1})}{\partial x_{1}}\frac{\partial p_{0}(x_{2})}{\partial x_{2}}\nonumber \\
  = &~ c_{1}P_{uu}+\varepsilon_{1}P_{up}+\frac{S_{upp}}{2}\varepsilon_{{\rm 0}}^{2},
\end{align}
where, for $i$, $j$, $k=u$, $p$,
\begin{align}
P_{ij} = &~ \int dxg_{i}^{1}\left(x\right)\int dx^{\prime}\frac{\partial F_{i}\left(x\right)}{\partial u_{j}\left(x^{\prime}\right)}u_{j}^{1}\left(x^{\prime}\right).\\
S_{ijk} = & ~ \int dxg_{i}^{1}\left(x\right)\int dx_{1}\int dx_{2}\frac{\partial^{2}F_{i}\left(x\right)}{\partial u_{j}\left(x_{1}\right)\partial u_{k}\left(x_{2}\right)}\nonumber \\
&\qquad\qquad\qquad\qquad\qquad\quad\frac{\partial u_{j}^{0}\left(x_{1}\right)}{\partial x_{1}}\frac{\partial u_{k}^{0}\left(x_{2}\right)}{\partial x_{2}}.
\end{align}
Hence we obtain 
\begin{equation}
J_{u}\frac{\partial c_{1}}{\partial t}=P_{uu}c_{1}+P_{up}\varepsilon_{1}+\frac{S_{upp}}{2}\varepsilon_{0}^{2}.\label{eq:c1_speed}
\end{equation}
Similarly, from Eq. (\ref{eq:depsilondcdt_b}), 
\begin{align}
I_{p}\frac{\partial\varepsilon_{{\rm 0}}}{\partial t}-I_{p}v_{\mathrm{nat}}-M_{p}v_{\mathrm{nat}}\varepsilon_{1} = &~ Q_{pp}\varepsilon_{0}+T_{ppu}\varepsilon_{0}c_{1}\nonumber\\
&+T_{ppp}\varepsilon_{0}\varepsilon_{1}+\frac{Q_{pppp}}{6}\varepsilon_{0}^{3}.\label{eq:e0_speed}\\
J_{p}\frac{\partial\varepsilon_{1}}{\partial t}-K_{p}v_{\mathrm{nat}}\varepsilon_{{\rm 0}} = & P_{pu}c_{1}+P_{pp}\varepsilon_{1}+\frac{S_{ppp}}{2}\varepsilon_{0}^{2},\label{eq:e1_speed}
\end{align}
where, for $i=u$, $p$,
\begin{equation}
K_{i}=\int dxg_{i}^{1}(x)\frac{\partial^{2}u_{i}^{0}(x)}{\partial x^{2}}. \label{eq:K_i}
\end{equation}

Since the solution to the above equations will be tedious, it is instructive
to interpret the equations from a symmetry point of view. This is
because when there is a separation between the exposed and inhibitory
profiles in the moving bump, the displacement mode will be coupled
with other distortion modes that prevent the profile separation from
diverging. Consider the coupling with the most important symmetric
mode, which is the width mode for weak inhibition, and the height
mode for strong inhibition \cite{Fung2010}. Irrespective of the details
of these modes, we can summarize the steady state equations (\ref{eq:c0_speed})
and (\ref{eq:e0_speed}) as
\begin{align}
-I_{u}v_{{\rm nat}}-M_{u}v_{{\rm nat}}c_{1} = &~ Q_{up}\varepsilon_{0}\nonumber \\ 
&+R_{u}\left(\varepsilon_{0}^{2},v_{{\rm nat}}\varepsilon_{0}\right)\varepsilon_{0},\label{eq:new_c0_intrinsic}\\
-I_{p}v_{{\rm nat}}-M_{p}v_{{\rm nat}}\varepsilon_{1} = &~ Q_{pp}\varepsilon_{0} \nonumber \\
&+R_{p}\left(\varepsilon_{0}^{2},v_{{\rm nat}}\varepsilon_{0}\right)\varepsilon_{0}.\label{eq:new_e0_intrinsic}
\end{align}
In Eq. (\ref{eq:new_c0_intrinsic}), we interpret $R_{u}\varepsilon_{0}$
as the force acting on the displacement mode due to the coupling with
the symmetric modes. Since the modes are decoupled when $\varepsilon_{0}$
vanishes, we consider forces proportional to $\varepsilon_{0}$. The
magnitudes of $R_{u}$ and $R_{p}$ depend on the following two factors.
(\ref{eq:dudt}) The distortions of the symmetric modes. Since the distortions
of the symmetric modes should be the same for $+\varepsilon_{0}$
and $-\varepsilon_{0}$, they should be proportional to $\varepsilon_{0}^{2}$.
(\ref{eq:dpdt}) It should depend on the bump velocity via $v_{{\rm nat}}\varepsilon_{0}$,
which originates from the wave terms of the moving symmetric mode.

Similarly, in the wave terms, $c_{1}$ and $\varepsilon_{1}$ can
be expressed as a linear combination of $\varepsilon_{0}^{2}$ and
$v_{{\rm nat}}\varepsilon_{0}$. Hence we can write 
\begin{align}
&-I_{u}v_{{\rm nat}}-M_{u1}v_{{\rm nat}}\varepsilon_{0}^{2}-M_{u2}v_{{\rm nat}}^{2}\varepsilon_{0} \nonumber \\
&\qquad\qquad\qquad =  Q_{up}\varepsilon_{0}+R_{u1}\varepsilon_{0}^{3}+R_{u2}v_{{\rm nat}}\varepsilon_{0}^{2},\label{eq:new_c0_intrinsic_2}\\
&-I_{p}v_{{\rm nat}}-M_{p1}v_{{\rm nat}}\varepsilon_{0}^{2}-M_{p2}v_{{\rm nat}}^{2}\varepsilon_{0} \nonumber \\
&\qquad\qquad\qquad =  Q_{pp}\varepsilon_{0}+R_{p1}\varepsilon_{0}^{3}+R_{p2}v_{{\rm nat}}\varepsilon_{0}^{2}.\label{eq:new_e0_intrinsic_2}
\end{align}
After elimination the variables $c_{1}$ and $\varepsilon_{1}$ using
Eqs. (\ref{eq:c1_speed}) and (\ref{eq:e1_speed}), we obtain
\begin{align}
R_{u1}  = &~ \frac{T_{upu}\left(P_{up}S_{ppp}-P_{pp}S_{upp}\right)}{2\left(P_{uu}P_{pp}-P_{pu}P_{up}\right)}\nonumber\\
&+\frac{T_{upp}\left(P_{pu}S_{upp}-P_{uu}S_{ppp}\right)}{2\left(P_{uu}P_{pp}-P_{pu}P_{up}\right)}+\frac{Q_{uppp}}{6},\label{eq:ru1}\\
R_{u2}  = &~ \frac{T_{upu}P_{up}K_{p}}{P_{uu}P_{pp}-P_{pu}P_{up}}-\frac{T_{upp}P_{uu}K_{p}}{P_{uu}P_{pp}-P_{pu}P_{up}},\\
R_{p1}  = &~ \frac{T_{ppu}\left(P_{up}S_{ppp}-P_{pp}S_{upp}\right)}{2\left(P_{uu}P_{pp}-P_{pu}P_{up}\right)}+\nonumber\\
&\frac{T_{ppp}\left(P_{pu}S_{upp}-P_{uu}S_{ppp}\right)}{2\left(P_{uu}P_{pp}-P_{pu}P_{up}\right)}+\frac{Q_{pppp}}{6},\\
R_{p2}  = &~ \frac{T_{ppu}P_{up}K_{p}}{P_{uu}P_{pp}-P_{pu}P_{up}}-\frac{T_{ppp}P_{uu}K_{p}}{P_{uu}P_{pp}-P_{pu}P_{up}},\\
M_{u1}  = &~ \frac{M_{u}\left(P_{up}S_{ppp}-P_{pp}S_{upp}\right)}{2\left(P_{uu}P_{pp}-P_{pu}P_{up}\right)},\\
M_{u2}  = &~ \frac{M_{u}P_{up}K_{p}}{P_{uu}P_{pp}-P_{pu}P_{up}},\\
M_{p1}  = &~ \frac{M_{p}\left(P_{pu}S_{upp}-P_{uu}S_{ppp}\right)}{2\left(P_{uu}P_{pp}-P_{pu}P_{up}\right)},\\
M_{p2}  = &~ -\frac{M_{p}P_{uu}K_{p}}{P_{uu}P_{pp}-P_{pu}P_{up}}.\label{eq:mp2}
\end{align}

In fact, the symmetric modes in Eqs. (\ref{eq:new_c0_intrinsic_2})
and (\ref{eq:new_e0_intrinsic_2}) may consist of more than one or
even all of them. We note that the relaxation rate eigenvalues do
not enter the equation here. From Eqs. (\ref{eq:new_c0_intrinsic_2})
and (\ref{eq:new_e0_intrinsic_2}),
\begin{align}
-v_{{\rm nat}} = &~ \frac{Q_{up}\varepsilon_{0}}{I_{u}}+\frac{R_{u1}\varepsilon_{0}^{3}}{I_{u}}+\frac{R_{u2}v_{{\rm nat}}\varepsilon_{0}^{2}}{I_{u}}\nonumber\\
&\qquad\qquad+\frac{M_{u1}v_{{\rm nat}}\varepsilon_{0}^{2}}{I_{u}}+\frac{M_{u2}v_{{\rm nat}}^{2}\varepsilon_{0}}{I_{u}},\\
-v_{{\rm nat}} = &~ \frac{Q_{pp}\varepsilon_{0}}{I_{p}}+\frac{R_{p1}\varepsilon_{0}^{3}}{I_{p}}+\frac{R_{p2}v_{{\rm nat}}\varepsilon_{0}^{2}}{I_{p}}\nonumber\\
&\qquad\qquad+\frac{M_{p1}v_{{\rm nat}}\varepsilon_{0}^{2}}{I_{p}}+\frac{M_{p2}v_{{\rm nat}}^{2}\varepsilon_{0}}{I_{p}}.
\end{align}
Note that $Q_{uu}+Q_{up}=0$ due to translational invariance. Equating
the two expressions of $v_{{\rm nat}}$, we arrive at an expression
for $\varepsilon_{{\rm int}}$,

\begin{align}
&\left(\frac{Q_{uu}}{I_{u}}+\frac{Q_{pp}}{I_{p}}\right)\varepsilon_{{\rm int}}\nonumber\\
= &~ \left(\frac{R_{u1}}{I_{u}}-\frac{R{}_{p1}}{I_{p}}\right)\varepsilon_{{\rm int}}^{3}\nonumber\\
& \qquad+\left(\frac{R_{u2}+M_{u1}}{I_{u}}-\frac{R{}_{p2}+M_{p2}}{I_{p}}\right)v_{{\rm nat}}\varepsilon_{{\rm int}}^{2}\nonumber \\
 & \qquad+\left(\frac{M_{u2}}{I_{u}}-\frac{M_{p2}}{I_{p}}\right)v_{{\rm nat}}^{2}\varepsilon_{{\rm int}}.\label{eq:eint}
\end{align}
Furthermore, from Eq. (\ref{eq:e0_speed}), we have, to the lowest
order,
\begin{equation}
\varepsilon_{{\rm int}}\approx v_{{\rm nat}}\tau_{{\rm int}},\quad\tau_{\mathrm{int}}\equiv-\frac{I_{p}}{Q_{pp}}.\label{eq:eint_v}
\end{equation}
$\tau_{\mathrm{int}}$ is an intrinsic time scale of the neural system.
Since $\varepsilon_{{\rm int}}$ is the lag of the inhibitory profile
relative to the exposed profile, it has the same sign as $v_{{\rm nat}}$.
This implies that $\tau_{\mathrm{int}}$ is positive. (Eq. (\ref{eq:c0_speed})
yields the same result if we make use of the trapnslational symmetry
relation $Q_{uu}+Q_{up}=0$ and note that $Q_{uu}/I_{u}+Q_{pp}/I_{p}\approx0$
near the critical point.) Introducing $K_{1}\equiv R_{u1}/I_{u}-R_{p1}/I_{p}$,
$K_{2}\equiv(R_{u2}+M_{u1})/I_{u}-(R_{p2}+M_{p1})/I_{p}$, $K_{3}\equiv M{}_{u2}/I_{u}-M_{p2}/I_{p}$,
we can express $v_{{\rm nat}}$ in terms of the eigenvalue in Eq.
(\ref{eq:eigenvalue}),
\begin{equation}
v_{{\rm nat}}=\pm\sqrt{\frac{\lambda}{K}},\label{eq:vnat_eigenvalue}
\end{equation}
where
\begin{equation}
K=K_{1}\tau_{{\rm int}}^{2}+K_{2}\tau_{{\rm int}}+K_{3}.\label{eq:k}
\end{equation}

In the static phase, $\lambda<0,$ and both $v_{\mathrm{nat}}$ and
$\varepsilon_{{\rm int}}$ vanish. In the moving phase, $\lambda>0,$
and the critical regime is given by $v_{\mathrm{nat}}\sim\varepsilon_{{\rm int}}\sim\sqrt{\lambda}$.

\section{Extrinsic Behavior\label{sec:tracking}}

Here we consider the network response to an external stimulus moving
with velocity $v_{I}$. The dynamical equations are analogous to those
in the previous section, except that an external stimulus is present
in the dynamical equation for the exposed profile, and the natural
velocity is replaced by the stimulus velocity $v_{I}$. 
\begin{align}
&\frac{\partial}{\partial t}\delta u\left(x\right)-v_{I}\frac{\partial u_{0}\left(x\right)}{\partial x} \nonumber\\
= &~\int dx^{\prime}\frac{\partial F_{u}\left(x\right)}{\partial u\left(x^{\prime}\right)}\delta u\left(x^{\prime}\right)+\int dx^{\prime}\frac{\partial F_{u}\left(x\right)}{\partial p\left(x^{\prime}\right)}\delta p\left(x^{\prime}\right)\nonumber\\
&+I^{{\rm ext}}\left(x\right),\label{eq:ddeltaudt_tracking}\\
&\frac{\partial}{\partial t}\delta p\left(x\right)-v_{I}\frac{\partial p_{0}\left(x\right)}{\partial x} \nonumber\\
= &~ \int dx^{\prime}\frac{\partial F_{p}\left(x\right)}{\partial u\left(x^{\prime}\right)}\delta u\left(x^{\prime}\right)+\int dx^{\prime}\frac{\partial F_{p}\left(x\right)}{\partial p\left(x^{\prime}\right)}\delta p\left(x^{\prime}\right).\label{eq:ddeltapdt_tracking}
\end{align}
Here, $x$ is the coordinate relative to the moving bump. Now we consider
the distortion due to the bump movement in the reference frame that
$c_{0}=0$, 
\begin{equation}
\delta u\left(x\right)=c_{1}u_{1}\left(x\right)\text{{\rm , }}\delta p\left(x\right)=\varepsilon_{0}\frac{\partial p_{0}}{\partial x}+\varepsilon_{1}p_{1}\left(x\right).
\end{equation}
To make the discussion more concrete, we consider stimuli having the
same profile as the bump, and the bump is displaced by $s$ relative
to the stimulus, that is,

\begin{equation}
I^{\mathrm{ext}}(x)=\frac{u_{0}(x+s)}{\tau_{\mathrm{stim}}}\approx\frac{1}{\tau_{\mathrm{stim}}}\left[u_{0}(x)+s\frac{\partial u_{0}(x)}{\partial x}\right],
\end{equation}
where the amplitude of the stimulus is given by the amplitude of $u_{0}(x)$
divided by $\tau_{\mathrm{stim}}$, referred to as the stimulus time.
While this definition is convenient for analytical purpose, in simulations
we use 
\begin{equation}
I^{\mathrm{ext}}(x)=\frac{A}{\tau_{s}}\exp\left[-\frac{\left(x-z_{I}\right)^{2}}{4a^{2}}\right].
\end{equation}
The corresponding $\tau_{{\rm stim}}$ can be approximated by $\max_{x}u_{0}\left(x\right)\tau_{s}/A$.
To reduce the numerical sensitivity to $\tilde{k}$, we further define
$\hat{A}\equiv\rho J_{0}A/\tilde{u}_{{\rm int}}$ where $\tilde{u}_{{\rm int}}\equiv\sqrt{8}(1+\sqrt{1-\tilde{k}})/\tilde{k}$
is the bump amplitude in the absence of external stimuli.

Multiplying both sides of Eq. (\ref{eq:ddeltaudt_tracking}) by $g_{u}^{0}$
and integrating, the last term in Eq. (\ref{eq:ddeltaudt_tracking})
becomes proportional to the displacement $s$. Following steps similar
to those in the previous section, we obtain the following equations
\begin{align}
-I_{u}v_{I}-M_{u}v_{I}c_{1} = &~ Q_{up}\varepsilon_{0}+T_{upu}\varepsilon_{0}c_{1}+T_{upp}\varepsilon_{0}\varepsilon_{1}\nonumber\\
&+\frac{Q_{uppp}}{6}\varepsilon_{0}^{3}+\frac{I_{u}}{\tau_{\mathrm{stim}}}s.\label{eq:c0_tracking}\\
0 = &~ P_{uu}c_{1}+P_{up}\varepsilon_{1}+\frac{S_{upp}}{2}\varepsilon_{0}^{2}+\frac{L_{u}}{\tau_{\mathrm{stim}}}.\label{eq:c1_tracking}\\
-I_{p}v_{I}-M_{p}v_{I}\varepsilon_{1} = &~ Q_{pp}\varepsilon_{0}+T_{ppu}\varepsilon_{0}c_{1}+T_{ppp}\varepsilon_{0}\varepsilon_{1}\nonumber\\
&+\frac{Q_{pppp}}{6}\varepsilon_{0}^{3}.\label{eq:e0_tracking}\\
-K_{p}v_{I}\varepsilon_{0} = &~ P_{pu}c_{1}+P_{pp}\varepsilon_{1}+\frac{S_{ppp}}{2}\varepsilon_{0}^{2}.\label{eq:e1_tracking}
\end{align}
In Eq. (\ref{eq:c1_tracking}), we have introduced
\begin{equation}
L_{u}=\int dxg_{u}^{1}(x)u_{0}(x).
\end{equation}

Interpreting the equations as those describing the dynamics coupled
to the symmetric modes, we can write 
\begin{align}
-I_{u}v_{I}-M_{u}v_{I}c_{1} = &~ Q_{up}\varepsilon_{0}+R_{u}\left(\varepsilon_{0}^{2},v_{I}\varepsilon_{0},\tau_{{\rm stim}}^{-1}\right)\varepsilon_{0}\nonumber \\
&+\frac{sI_{u}}{\tau_{{\rm stim}}},\label{eq:new_c0_tracking}\\
-I_{p}v_{I}-M_{p}v_{I}\varepsilon_{1} = &~ Q_{pp}\varepsilon_{0}+R_{p}\left(\varepsilon_{0}^{2},v_{I}\varepsilon_{0},\tau_{{\rm stim}}^{-1}\right)\varepsilon_{0}.\label{eq:new_e0_tracking}
\end{align}
The interpretation of $R_{u}\varepsilon_{0}$ is the same as that
in Eq. (\ref{eq:new_c0_intrinsic}), except that the force acting
on the displacement mode has an additional dependence on the distortion
of the symmetric modes directly due to the external stimulus. Hence
we have introduced the third argument of $\tau_{{\rm stim}}^{-1}$
in $R_{u}$. Similarly, in the wave terms, $c_{1}$ and $\varepsilon_{1}$
can be expressed as a linear combination of $\varepsilon_{0}^{2}$,
$v_{I}\varepsilon_{0}$ and, additionally, $\tau_{{\rm stim}}^{-1}$.
Hence we can write
\begin{align}
&-I_{u}v_{I}-M_{u1}v_{I}\varepsilon_{0}^{2}-M_{u2}v_{I}^{2}\varepsilon_{0}-\frac{M_{u3}v_{I}}{\tau_{{\rm stim}}} \nonumber\\
= &~Q_{up}\varepsilon_{0}+R_{u1}\varepsilon_{0}^{3}+R_{u2}v_{I}\varepsilon_{0}^{2}+\frac{R_{u3}\varepsilon_{0}}{\tau_{{\rm stim}}}+\frac{sI_{u}}{\tau_{{\rm stim}}},\nonumber \\
\label{eq:new_c0_tracking_2}\\
&-I_{p}v_{I}-M_{p1}v_{I}\varepsilon_{0}^{2}-M_{p2}v_{I}^{2}\varepsilon_{0}-\frac{M_{p3}v_{I}}{\tau_{{\rm stim}}}\nonumber\\ 
= &~ Q_{pp}\varepsilon_{0}+R_{p1}\varepsilon_{0}^{3}+R_{p2}v_{I}\varepsilon_{0}^{2}+\frac{R_{p3}\varepsilon_{0}}{\tau_{{\rm stim}}}.\label{eq:new_e0_tracking_2}
\end{align}
After eliminating the variables $c_{1}$ and $\varepsilon_{1}$ from
their dynamical equations, we can derive expressions of $R_{u1}$,
$R_{u2}$, $R_{p1}$, $R_{p2}$, $M_{u1}$, $M_{u2}$, $M_{p1}$,
$M_{p2}$ identical to Eqs. (\ref{eq:ru1}) to (\ref{eq:mp2}). In
addition,
\begin{align}
R_{u3} & =  \frac{T_{upu}P_{pp}L_{u}}{P_{uu}P_{pp}-P_{pu}P_{up}}+\frac{T_{upp}P_{pu}L_{u}}{P_{uu}P_{pp}-P_{pu}P_{up}},\\
R_{p3} & =  \frac{T_{ppu}P_{pp}L{}_{u}}{P_{uu}P_{pp}-P_{pu}P_{up}}+\frac{T_{ppp}P_{pu}L{}_{u}}{P_{uu}P_{pp}-P_{pu}P_{up}},\\
M_{u3} & =  -\frac{M_{u}P_{pp}L_{u}}{P_{uu}P_{pp}-P_{pu}P_{up}},\\
M_{p3} & =  -\frac{M_{p}P_{pu}L_{u}}{P_{uu}P_{pp}-P_{pu}P_{up}}.
\end{align}
From Eqs. (\ref{eq:new_c0_tracking_2}) and (\ref{eq:new_e0_tracking_2}),
\begin{align}
-v_{I} = &~\frac{Q_{up}\varepsilon_{0}}{I_{u}}+\frac{R_{u1}\varepsilon_{0}^{3}}{I_{u}}+\frac{R_{u2}v_{I}\varepsilon_{0}^{2}}{I_{u}}+\frac{R_{u3}\varepsilon_{0}}{\tau_{{\rm stim}}I_{u}}\nonumber \\
&+\frac{M_{u1}v_{I}\varepsilon_{0}^{2}}{I_{u}}+\frac{M_{u2}v_{I}^{2}\varepsilon_{0}}{I_{u}}+\frac{M_{u3}v_{I}}{\tau_{{\rm stim}}I_{u}}+\frac{s}{\tau_{{\rm stim}}},\nonumber \\
\\
-v_{I} = &~ \frac{Q_{pp}\varepsilon_{0}}{I_{p}}+\frac{R_{p1}\varepsilon_{0}^{3}}{I_{p}}+\frac{R_{p2}v_{I}\varepsilon_{0}^{2}}{I_{p}}+\frac{R_{p3}\varepsilon_{0}}{\tau_{{\rm stim}}I_{p}}\nonumber\\
&+\frac{M_{p1}v_{I}\varepsilon_{0}^{2}}{I_{p}}+\frac{M_{p2}v_{I}^{2}\varepsilon_{0}}{I_{p}}+\frac{M_{p3}v_{I}}{\tau_{{\rm stim}}I_{p}}.
\end{align}
Note that $Q_{uu}+Q_{up}=0$ due to translational invariance. Eliminating
$v_{I}$, 
\begin{align}
&\left(\frac{Q_{uu}}{I_{u}}+\frac{Q_{pp}}{I_{p}}\right)\varepsilon_{0}-\left(\frac{R_{u1}}{I_{u}}-\frac{R_{p1}}{I_{p}}\right)\varepsilon_{0}^{3}\nonumber\\
&-\left(\frac{R_{u2}+M_{u1}}{I_{u}}-\frac{R_{p2}+M_{p1}}{I_{p}}\right)v_{I}\varepsilon_{0}^{2}\nonumber \\
&-\left(\frac{R_{u3}}{I_{u}}-\frac{R_{p3}}{I_{p}}\right)\frac{\varepsilon_{0}}{\tau_{{\rm stim}}}\nonumber\\
&-\left(\frac{M_{u2}}{I_{u}}-\frac{M_{p2}}{I_{p}}\right)v_{I}^{2}\varepsilon_{0}-\left(\frac{M_{u3}}{I_{u}}-\frac{M_{p3}}{I_{p}}\right)\frac{v_{I}}{\tau_{{\rm stim}}} \nonumber\\ 
= &~ \frac{s}{\tau_{{\rm stim}}}.
\end{align}
Recall that the instability eigenvalue is given by $\lambda=Q_{uu}/I_{u}+Q_{pp}/I_{p}$.
Besides the definitions of $K_{1}$, $K_{2}$ and $K_{3}$, we further
introduce $K_{4}\equiv R_{u3}/I_{u}-R_{p3}/I_{p}$, $K_{5}\equiv M_{u3}/I_{u}-M_{p3}/I_{p}$.
Then we have
\begin{align}
\lambda\varepsilon_{0}-K_{1}\varepsilon_{0}^{3}-K_{2}v_{I}\varepsilon_{0}^{2}-K_{3}v_{I}^{2}\varepsilon_{0}&\nonumber\\
-K_{4}\frac{\varepsilon_{0}}{\tau_{{\rm stim}}}-K_{5}\frac{v_{I}}{\tau_{{\rm stim}}}&=\frac{s}{\tau_{{\rm stim}}}.\label{eq:lambda_epsilon_0}
\end{align}
Let us compare this equation with the case of the bump's intrinsic
motion. The latter case can be done by replacing $v_{I}$ with $v_{{\rm nat}}$,
$\varepsilon_{0}$ by $\varepsilon_{{\rm int}}$ and $\tau_{{\rm stim}}^{-1}=0$,
as verified in Eq. (\ref{eq:eint}). This leads to
\begin{equation}
\lambda\varepsilon_{{\rm int}}-K_{1}\varepsilon_{{\rm int}}^{3}-K_{2}v_{{\rm nat}}\varepsilon_{{\rm int}}^{2}-K_{3}v_{{\rm nat}}^{2}\varepsilon_{{\rm int}}=0.\label{eq:eq:lambda_epsilon_int}
\end{equation}
For the lowest order terms in Eq. (\ref{eq:new_e0_tracking_2}), we
obtain
\begin{equation}
\varepsilon_{0}=v_{I}\tau_{{\rm int}},
\end{equation}
similar to Eq. (\ref{eq:eint_v}) for the intrinsic motion. The anticipation
time is defined by
\begin{equation}
\tau_{{\rm ant}}=\frac{s}{v_{I}}.\label{eq:app_tau_ant}
\end{equation}
Substituting Eqs. (\ref{eq:eq:lambda_epsilon_int}) - (\ref{eq:app_tau_ant})
into Eq. (\ref{eq:lambda_epsilon_0}), and introducing $\tau_{{\rm con}}=-K_{4}\varepsilon_{{\rm int}}-K_{5}$,
we arrive at,
\begin{equation}
\tau_{{\rm ant}}=K\tau_{{\rm stim}}\tau_{{\rm int}}\left(v_{{\rm nat}}^{2}-v_{I}^{2}\right)+\tau_{{\rm con}}.\label{eq:app_tau_ant_general}
\end{equation}
In the limit of weak and slowly moving stimulus, in which $\tau_{\mathrm{stim}}$
is large and $v_{I}$ is small, the anticipation time reduces to the
transparent form
\begin{equation}
\tau_{\mathrm{ant}}=\tau_{\mathrm{stim}}\tau_{\mathrm{int}}\lambda.\label{eq:app_tau_anticipation_slow}
\end{equation}

\section{Response to Noises \label{sec:noise}}

From the viewpoint of fluctuation-response relations, we would like
to connect our results with thermal fluctuations. Hence we consider
the dynamics in the presence of thermal noises by modifying Eq. (\ref{eq:dudt}),
\begin{equation}
\frac{\partial u\left(x\right)}{\partial t}=F_{u}\left[x;u,p\right]-\eta\left(t\right)\frac{\partial u_{0}}{\partial x}\text{{\rm , }}\quad\frac{\partial p\left(x\right)}{\partial t}=F_{p}\left[x;u,p\right],\label{eq:dudtdpdt_resp_to_noise}
\end{equation}
where
\begin{equation}
\left\langle \eta\left(t\right)\right\rangle =0\text{{\rm , and }}\left\langle \eta\left(t\right)\eta\left(t^{\prime}\right)\right\rangle =2T\delta\left(t-t^{\prime}\right).\label{eq:eta_variance_resp_to_noise}
\end{equation}
We first consider the static phase. Eq. (\ref{eq:depsilondcdt_a}) implies that
\begin{align}
\frac{\partial}{\partial t}\delta u\left(x\right)=&~\int dx^{\prime}\frac{\partial F_{u}\left(x\right)}{\partial u\left(x^{\prime}\right)}\delta u\left(x^{\prime}\right)+\int dx^{\prime}\frac{\partial F_{u}\left(x\right)}{\partial p\left(x^{\prime}\right)}\delta p\left(x^{\prime}\right)\nonumber\\
&-\eta\left(t\right)\frac{\partial u_{0}}{\partial x}.
\end{align}

Following the analysis in Sec. \ref{sec:separation}, 
we arrive at 
\begin{equation}
\frac{\partial}{\partial t}\left(\begin{array}{c}
c_{0}\\
\varepsilon_{0}
\end{array}\right)=\left(\begin{array}{cc}
Q_{uu}/I_{u} & -Q_{uu}/I_{u}\\
-Q_{pp}/I_{p} & Q_{pp}/I_{p}
\end{array}\right)\left(\begin{array}{c}
c_{0}\\
\varepsilon_{0}
\end{array}\right)-\left(\begin{array}{c}
\eta\left(t\right)\\
0
\end{array}\right).
\end{equation}
This implies that 
\begin{equation}
\frac{\partial}{\partial t}\left(\varepsilon_{0}-c_{0}\right)=\lambda\left(\varepsilon_{0}-c_{0}\right)+\eta\left(t\right).
\end{equation}
The solution to this differential equation is 
\begin{equation}
\varepsilon_{0}-c_{0}=\int_{-\infty}^{t}dt^{\prime}\exp\left[\lambda\left(t-t^{\prime}\right)\right]\eta\left(t^{\prime}\right).
\end{equation}
Averaging over thermal noises, $\left\langle \varepsilon_{0}-c_{0}\right\rangle =0$
and
\begin{align}
&\left\langle \left(\varepsilon_{0}-c_{0}\right)^{2}\right\rangle \nonumber\\
=&\int_{-\infty}^{t}dt_{1}\int_{-\infty}^{t}dt_{2} e^{ \lambda\left[\left(t-t_{1}\right)+\left(t-t_{2}\right)\right]} \left\langle \eta\left(t_{1}\right)\eta\left(t_{2}\right)\right\rangle .
\end{align}
Using the noise average in Eq. (\ref{eq:eta_variance_resp_to_noise}),
\begin{equation}
\left\langle \left(\varepsilon_{0}-c_{0}\right)^{2}\right\rangle =2T\int_{-\infty}^{t}dt^{\prime}\exp\left[2\lambda\left(t-t^{\prime}\right)\right]=-\frac{T}{\lambda}.
\end{equation}
Equation (\ref{eq:app_tau_ant_general}) can now be cast into the form
of a fluctuation response relation. In this case, the response term
is the effective anticipation rate, that is, the inverse of the anticipation
time minus its value at the confluence point, 
\begin{equation}
\frac{\left\langle \left(\varepsilon_{0}-c_{0}\right)^{2}\right\rangle }{T}=-\frac{\tau_{\mathrm{stim}}\tau_{\mathrm{int}}}{\tau_{\mathrm{ant}}-\tau_{\mathrm{con}}}.
\end{equation}
This shows that the effective anticipation time in the static phase
is negative. The relation means that when the fluctuations of the
separation between the exposed and inhibitory profiles have a faster
rate of increase with the noise temperature, the network becomes more
responsive to the moving stimulus by shortening the delay time. At
the boundary of the static phase, fluctuations diverge and the bump
is in a ready-to-go state.

\renewcommand{\thefigure}{F.\arabic{figure}}
\setcounter{figure}{0}

\begin{figure*}
\begin{centering}
\includegraphics[width=\textwidth]{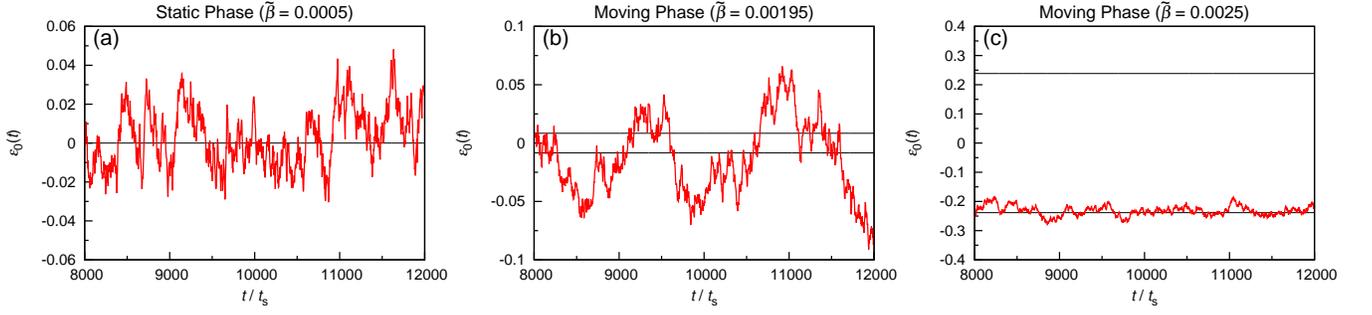}
\par\end{centering}

\protect\caption{\label{fig:fluctuation} (color online) Three samples of $\varepsilon_{0}\left(t\right)$
with dfferent values of $\tilde{\beta}$. (a) $\varepsilon_{0}\left(t\right)$
in the static phase. (b) $\varepsilon_{0}\left(t\right)$ in the moving
phase near the static-moving transition. (c) $\varepsilon_{0}\left(t\right)$
in the moving phase. Parameters: $\tilde{k}=0.3$, $\tau_{{\rm d}}=50\tau_{{\rm s}}$
and $T=1\times10^{-6}$.}

\end{figure*}

Next, we consider the behavior in the moving phase. We start with
the dynamical equations in the moving phase and in the presence of
an external stimulus. We consider the case that the dynamics is dominated
by a relaxation rate of the order $\lambda$, which is much slower
than those of other distortion modes. For the example of SFA, we see
that after the exposed profile couples with the inhibitory profile
with a slow relaxation rate $\tau_{i}^{-1}$, there exists a family
of inhibitory-like modes with relaxation rates approximately $\tau_{i}^{-1}$.
Hence, we consider the regime $\lambda\ll\tau_{i}^{-1}$. (We conjecture
that even when this condition is not satisfied, our analysis is still
applicable because the inhibitory-like modes are weakly coupled with
the external environment. We will leave this for further investigation.)
This implies that the symmetric modes are effectively remaining at
the instantaneous steady state. Hence interpreting the forces on the
displacement modes as the couplings with the symmetric modes, we rewrite
Eqs. (\ref{eq:new_c0_tracking_2}) and (\ref{eq:new_e0_tracking_2})
as 
\begin{align}
&-I_{u}v_{{\rm nat}}-M_{u1}v_{{\rm nat}}\varepsilon_{{\rm int}}^{2}-M_{u2}v_{{\rm nat}}^{2}\varepsilon_{{\rm int}} \nonumber \\
&\qquad= Q_{up}\varepsilon_{{\rm int}}+R_{u1}\varepsilon_{{\rm int}}^{3}+R_{u2}v_{{\rm nat}}\varepsilon_{{\rm int}}^{2}-I_{u}\eta,\\
&-I_{p}v_{{\rm nat}}-M_{p1}v_{{\rm nat}}\varepsilon_{{\rm int}}^{2}-M_{p2}v_{{\rm nat}}^{2}\varepsilon_{{\rm int}} \nonumber\\
&\qquad= Q_{pp}\varepsilon_{{\rm int}}+R_{p1}\varepsilon_{{\rm int}}^{3}+R_{p2}v_{{\rm nat}}\varepsilon_{{\rm int}}^{2},
\end{align}
where $\eta$ is the positional noise defined in the main text. Considering
the fluctuations around $v_{{\rm nat}}$ and $\varepsilon_{{\rm int}}$,
\begin{align}
   & -I_{u}\delta v-M_{u1}\varepsilon_{{\rm int}}^{2}\delta v-2M_{u1}v_{{\rm nat}}\varepsilon_{{\rm int}}\delta\varepsilon_{0}\nonumber\\
   & -M_{u2}v_{{\rm nat}}^{2}\delta\varepsilon_{0}-2M_{u2}v_{{\rm nat}}\varepsilon_{{\rm int}}\delta v\nonumber \\
  &\qquad\qquad = Q_{up}\delta\varepsilon_{0}+3R_{u1}\varepsilon_{{\rm int}}^{2}\delta\varepsilon_{0}+R_{u2}\varepsilon_{{\rm int}}^{2}\delta v\nonumber\\
  &\qquad\qquad ~~~~ +2R_{u2}v_{{\rm nat}}\varepsilon_{{\rm int}}\delta\varepsilon_{0}-I_{u}\eta,\\
   & I_{p}\frac{d}{dt}\delta\varepsilon_{0}-I_{p}\delta v-M_{p1}\varepsilon_{{\rm int}}^{2}\delta v-2M_{p1}v_{{\rm nat}}\varepsilon_{{\rm int}}\delta\varepsilon_{0}\nonumber\\
   &-M_{p2}v_{{\rm nat}}^{2}\delta\varepsilon_{0}-2M_{p2}v_{{\rm nat}}\varepsilon_{{\rm int}}\delta v\nonumber \\
  &\qquad\qquad=  Q_{pp}\delta\varepsilon_{0}+3R_{p1}\varepsilon_{{\rm int}}^{2}\delta\varepsilon_{0}+R_{p2}\varepsilon_{{\rm int}}^{2}\delta v\nonumber\\
  &\qquad\qquad ~~~~ +2R_{p2}v_{{\rm nat}}\varepsilon_{{\rm int}}\delta\varepsilon_{0}.
\end{align}
Eliminating $\delta v$,
\begin{align}
&\frac{d}{dt}\delta\varepsilon_{0} \nonumber\\
 = &~ \left(\frac{Q_{uu}}{I_{u}}+\frac{Q_{pp}}{I_{p}}\right)\delta\varepsilon_{0}-3\left(\frac{R_{u1}}{I_{u}}-\frac{R_{p1}}{I_{p}}\right)\varepsilon_{{\rm int}}^{2}\delta\varepsilon_{0}\nonumber\\
 &-\left(\frac{R_{u2}+M_{u1}}{I_{u}}-\frac{R_{p2}+M_{p1}}{I_{p}}\right)\varepsilon_{{\rm int}}^{3}\delta v\nonumber \\
   & -2\left(\frac{R_{u2}+M_{u1}}{I_{u}}-\frac{R_{p2}+M_{p1}}{I_{p}}\right)v_{{\rm nat}}\varepsilon_{{\rm int}}\delta\varepsilon_{0}\nonumber\\
   & -2\left(\frac{M_{u2}}{I_{u}}-\frac{M_{p2}}{I_{p}}\right)v_{{\rm nat}}\varepsilon_{{\rm int}}\delta v\nonumber \\
   & -\left(\frac{M_{u2}}{I_{u}}-\frac{M_{p2}}{I_{p}}\right)v_{{\rm nat}}^{2}\delta\varepsilon_{0}+\eta\nonumber \\
  = &~ \lambda\delta\varepsilon_{0}-3K_{1}\varepsilon_{{\rm int}}^{2}\delta\varepsilon_{0}-K_{2}\varepsilon_{{\rm int}}^{2}\delta v-K_{2}v_{{\rm nat}}\varepsilon_{{\rm int}}\delta\varepsilon_{0}\nonumber\\
  &-K_{4}v_{{\rm nat}}\varepsilon_{{\rm int}}\delta v-K_{4}v_{{\rm nat}}^{2}\delta\varepsilon_{0}+\eta\nonumber \\
\end{align}
Using Eq. (\ref{eq:eq:lambda_epsilon_int}) to eliminate $\lambda$,
and $\delta\varepsilon_{0}=\tau_{{\rm int}}\delta v$,
\begin{equation}
\frac{d}{dt}\delta\varepsilon_{0}=-2\lambda\delta\varepsilon_{0}+\eta.
\end{equation}
Solving the differential equation,
\begin{equation}
\delta\varepsilon_{0}\left(t\right)=\int_{-\infty}^{t}dt^{\prime}\exp\left[-2\lambda\left(t-t^{\prime}\right)\right]\eta\left(t\right).
\end{equation}
Fluctuations are given by
\begin{align}
&\left\langle \delta\varepsilon_{0}\left(t\right)^{2}\right\rangle  \nonumber\\
= &~ \int_{-\infty}^{t}dt_{2}\int_{-\infty}^{t}dt_{1}e^{-2\lambda\left(t-t_{1}\right)-2\lambda\left(t-t_{2}\right)}\left\langle \eta\left(t_{1}\right)\eta\left(t_{2}\right)\right\rangle \\
 = &~ \frac{T}{2\lambda}.
\end{align}
Connecting with the fluctuations with the response behavior through
Eq. (\ref{eq:app_tau_ant_general}),
\begin{align}
\frac{\left\langle \delta\varepsilon_{0}\left(t\right)^{2}\right\rangle }{T}&=\frac{\tau_{{\rm stim}}\tau_{{\rm int}}}{2\left(\tau_{{\rm ant}}-\tau_{{\rm con}}\right)}\left(\frac{v_{{\rm nat}}^{2}-v_{I}^{2}}{v_{{\rm nat}}^{2}}\right)\nonumber\\
&\xrightarrow{\left|v_{{\rm nat}}\right|\gg\left|v_{I}\right|}\frac{\tau_{{\rm stim}}\tau_{{\rm int}}}{2\left(\tau_{{\rm ant}}-\tau_{{\rm con}}\right)}
\end{align}

\section{Numerical Measurement of $\left\langle \delta\varepsilon_{0}^{2}\right\rangle$}\label{sec:numeric}

The variance of $\varepsilon_{0}\left(t\right)$ can be easily obtained
from simulations, if the set of parameters is chosen to be far from
phase boundaries. Those examples for CANNs with STD are shown in Fig.
\ref{fig:fluctuation} (a) and (c). In Fig. \ref{fig:fluctuation}(a),
$\tilde{\beta}$ is small enough to have a stable static fixed point
solution. In this case, there is only one fixed point solution of
$\varepsilon_{0}=0$. The statistics of $\varepsilon_{0}\left(t\right)$
is relatively simple. For a large enough $\tilde{\beta}$, as shown
in Fig. \ref{fig:fluctuation}(c), the two fixed point solutions to
$\varepsilon_{0}$ have opposite signs and are separated far apart.
As a result, $\varepsilon_{0}\left(t\right)$ will mostly stick to
one of the fixed point solution. The statistics of $\varepsilon_{0}\left(t\right)$
is similar to that of the static phase.

However, in the moving phase near the phase boundary, e.g. Fig. \ref{fig:fluctuation}(b),
the statistics may be problematic. The problem is due to the difference
between two fixed point solutions being too small, so that $\varepsilon_{0}\left(t\right)$
is fluctuating around two fixed point solutions ($\varepsilon_{0,{\rm fixed}}^{+}$
and $\varepsilon_{0,{\rm fixed}}^{-}$), even though the noise temperature
$T$ is small. Whenever $\varepsilon_{0}\left(t\right)$ is between
two fixed point solutions, attractions due to fixed point solutions
can affect our estimations of the variance of $\varepsilon_{0}\left(t\right)$
around a \textit{single} fixed point solution.

To overcome the interference between two fixed point solutions, a
trick is needed to filter out some data. In the statistics of Fig.
4 in the main text, we have discarded $\varepsilon_{0}\left(t\right)$
less than $\left|\varepsilon_{0,{\rm fixed}}^{+}\right|$. So, we
approximate the variance by
\begin{equation}
{\rm Var}\left[\varepsilon_{0}\left(t\right)-\varepsilon_{0,{\rm fixed}}^{\pm}\right]=\frac{\sum_{t^{\prime}\in S}\left[\left|\varepsilon_{0}\left(t^{\prime}\right)\right|-\left|\varepsilon_{0,{\rm fixed}}^{\pm}\right|\right]^{2}}{N_{{\rm sample}}-1},
\end{equation}
where $S\equiv\left\{ t^{\prime}\left|\left|\varepsilon_{0}\left(t^{\prime}\right)\right|>\left|\varepsilon_{0,{\rm fixed}}^{\pm}\right|\right.\right\} $
and $N_{{\rm sample}}\equiv\left|S\right|$.

\end{document}